\documentclass[%
 reprint,
superscriptaddress,
 amsmath,amssymb,
 aps,
]{revtex4-2}

\usepackage{graphicx}
\usepackage{dcolumn}
\usepackage{bm}
\usepackage{hyperref}
\usepackage[utf8]{inputenc}
\usepackage{fontenc}

\hypersetup{colorlinks=true, linkcolor=blue, citecolor=blue, filecolor=blue, urlcolor=blue}

\begin{document}


\title{Scaling Behavior of Magnetoresistance and Hall Resistivity in Altermagnet CrSb}
\author{Xin Peng}\thanks{These authors contribute equally.}
\affiliation{Department of Physics, China Jiliang University, Hangzhou 310018, China}
\author{Yuzhi Wang}\thanks{These authors contribute equally.}
\affiliation{Beijing National Laboratory for Condensed Matter Physics and Institute of physics, Chinese academy of sciences, Beijing 100190, China}
\affiliation{University of Chinese academy of sciences, Beijing 100049, China}

\author{Shengnan Zhang}
\affiliation{Beijing Polytechnic College, Beijing 100042, China}
\author{Yi Zhou}
\affiliation{Department of Physics, China Jiliang University, Hangzhou 310018, China}
\author{Yuran Sun}
\affiliation{Department of Physics, China Jiliang University, Hangzhou 310018, China}
\author{Yahui Su}
\affiliation{Department of Physics, China Jiliang University, Hangzhou 310018, China}
\author{Chunxiang Wu}
\affiliation{School of Physics, Zhejiang University, Hangzhou 310027, China}
\author{Tingyu Zhou}
\affiliation{School of Physics, Zhejiang University, Hangzhou 310027, China}
\author{Le Liu}
\affiliation{School of Physics, Zhejiang University, Hangzhou 310027, China}
\author{Hangdong Wang}
\affiliation{School of Physics, Hangzhou Normal University, Hangzhou 310036, China}
\author{Jinhu Yang}
\affiliation{School of Physics, Hangzhou Normal University, Hangzhou 310036, China}
\author{Bin Chen}
\affiliation{School of Physics, Hangzhou Normal University, Hangzhou 310036, China}
\author{Zhong Fang}
\affiliation{Beijing National Laboratory for Condensed Matter Physics and Institute of physics, Chinese academy of sciences, Beijing 100190, China}
\affiliation{University of Chinese academy of sciences, Beijing 100049, China}
\author{Jianhua Du}\email{Corresponding author:	jhdu@cjlu.edu.cn}
\affiliation{Department of Physics, China Jiliang University, Hangzhou 310018, China}
\author{Zhiwei Jiao}\email{jiaozw@cjlu.edu.cn}
\affiliation{Department of Physics, China Jiliang University, Hangzhou 310018, China}
\author{Quansheng Wu}\email{quansheng.wu@iphy.ac.cn}
\affiliation{Beijing National Laboratory for Condensed Matter Physics and Institute of physics, Chinese academy of sciences, Beijing 100190, China}
\affiliation{University of Chinese academy of sciences, Beijing 100049, China}
\author{Minghu Fang}\email{mhfang@zju.edu.cn}
\affiliation{School of Physics, Zhejiang University, Hangzhou 310027, China}
\affiliation{Collaborative Innovation Center of Advanced Microstructure, Nanjing University, Nanjing 210093, China}


%


\date{\today}

\begin{abstract}
The discovery of altermagnet (AM) marks a significant advancement in magnetic materials, combining characteristics of both ferromagnetism and antiferromagnetism. In this Letter, we focus on CrSb, which has been verified to be an AM and to exhibit substantial spin splitting near the Fermi level. After successfully growing high-quality CrSb single crystals, we performed comprehensive magnetization, magnetoresistance (MR), and Hall resistivity measurements, along with the electronic structure, and Fermi surface (FS) calculations, as well as the magneto-transport property numerical simulations. An antiferromagnetic transition occurring at $T_{N}$ = 712 K was reconfirmed. It was found that both experimental MR and Hall resistivity are consistent with the numerical simulation results, and exhibit obvious scaling behavior. The nonlinear Hall resistivity is due to its multi-band structure, rather than an anomalous Hall effect (AHE). Especially, the scaling behavior in Hall resistivity is first observed within an AM material. These findings demonstrate that the magneto-transport properties in CrSb originate from the intrinsic electronic structure and are dominated by the Lorentz force.         
\end{abstract}

                
\maketitle



The exploration of magnetic materials has advanced significantly with the recent identification of a novel state known as altermagnetism (AM) \cite{PhysRevX.12.040002,PhysRevX.12.031042, PhysRevX.12.040501}, which is characterized by a unique combination of antiferromagnetic and ferromagnetic properties, where time-reversal ($\mathcal{T}$), parity ($\mathcal{P}$), and lattice translation (t$\mathcal{T}$) symmetries are broken. As a result, AM materials exhibit spin-split electronic bands akin to ferromagnet, despite having zero net macroscopic magnetization. This intriguing behavior has been confirmed in several AM candidates by angle-resolved photoemission spectroscopy (ARPES) \cite{PhysRevLett.132.036702,PhysRevB.109.115102,krempasky2024altermagnetic,reimers2024direct,fedchenko2024} measurements, revealing phenomena such as the anomalous Hall effect (AHE) \cite{PhysRevLett.130.036702,wang2023emergent,feng2022anomalous,tschirner2023saturation} and spin current generation \cite{PhysRevLett.128.197202,bose2022tilted}, which have attracted considerable interest due to their promising application in spintronics. Among the numerous theoretically predicted AM candidates, RuO$_{2}$, MnTe, and MnTe$_{2}$ have been extensively studied~\cite{PhysRevX.12.040501,fedchenko2024,feng2022anomalous,tschirner2023saturation,wang2023emergent,bose2022tilted,PhysRevLett.128.197202,krempasky2024altermagnetic,PhysRevLett.132.036702,PhysRevMaterials.8.L041402,zhu2024observation}. However, CrSb has received less attention despite its notably high Néel temperature \cite{haraldsen1946system,PhysRev.85.365,PhysRev.129.2008} and substantial spin-splitting energy predicted to reach up to 1.2 eV near the Fermi level (\textit{E}$_{F}$) \cite{PhysRevX.12.031042,guo2023spin}. CrSb crystallizes in a hexagonal NiAs-type structure and features a collinear alignment of antiparallel spins along the \textit{c}-axis [see the inset of Fig. 1(a)], as determined by neutron diffraction measurements \cite{PhysRev.85.365}. This magnetic arrangement connects the spin sublattices through screw and mirror operations, classifying CrSb as a bulk \textit{g}-wave altermagnet \cite{PhysRevX.12.031042,PhysRevLett.133.206401}. The complex nature of its spin band splitting suggests the potential for unique behaviors in its transport properties. Despite these promising attributes, a detailed investigation of its transport properties has been lacking. Previous studies have hinted at interesting electronic features, yet high-resolution spectroscopic investigations are necessary to fully elucidate the nature of band splitting and the resulting transport phenomena. Therefore, there is an urgent need to investigate its magnetoresistance (MR) and Hall resistivity to evaluate its viability for spintronic applications. 

In this Letter, we successfully grew high-quality CrSb single crystals, and conducted magnetization measurements, reconfirming an  antiferromagnetic transition occurring at $T_{N}$ = 712 K. By combining the electronic structure, Fermi surface (FS) calculations, as well as the magneto-transport numerical simulations, we studied the scaling behavior of longitudinal magnetoresistance (MR) and Hall resistivity. We found that the electronic transport is determined by its intrinsic electronic structure, even in this AM material, especially, the nonlinear Hall resistivity results from its multi-band structure rather than being an anomalous Hall effect. These results indicate a challenge in spintronic applications by using AM CrSb.  


\begin{figure}[!htbpb]
	\includegraphics[width= 8.6cm]{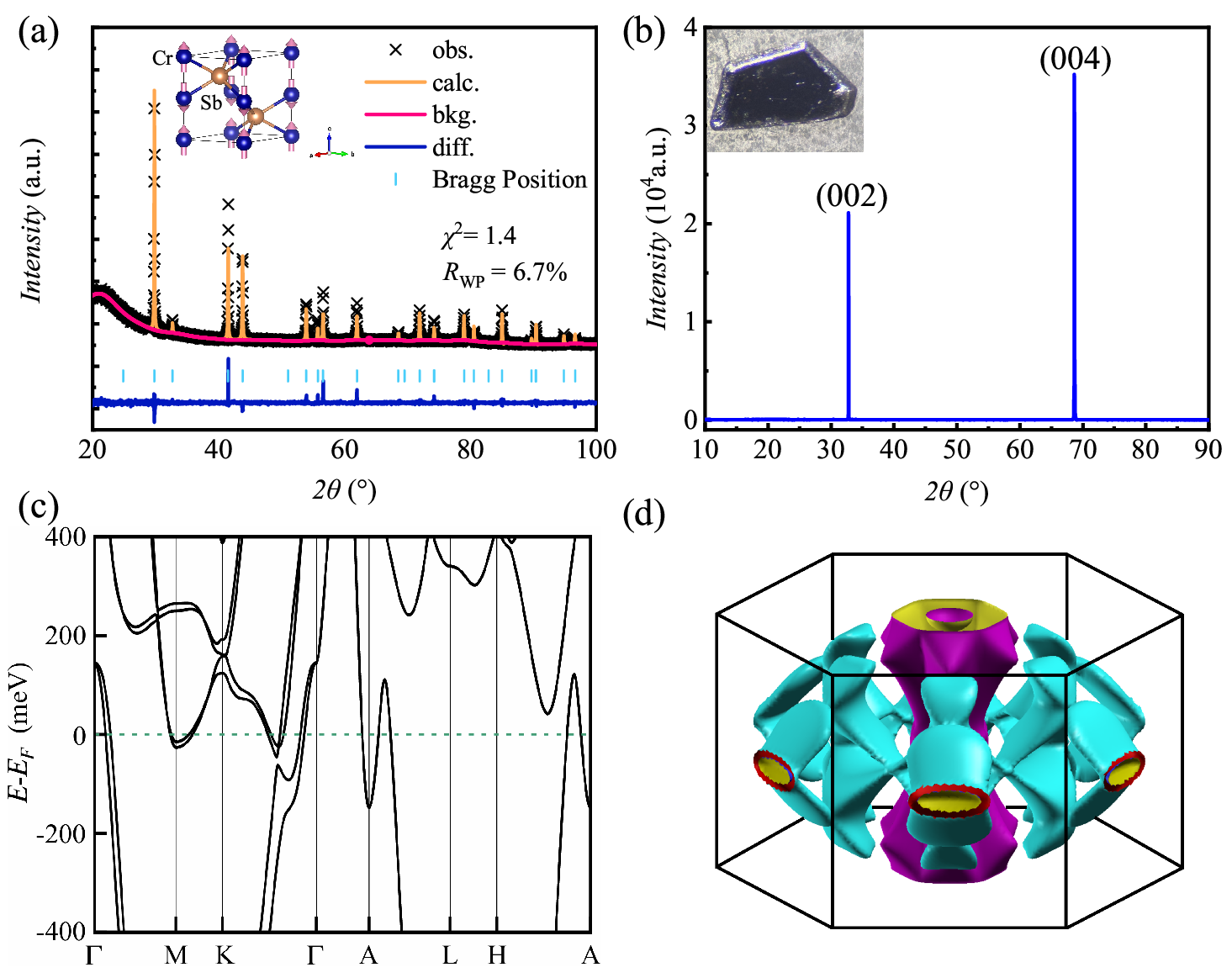}
	\caption{\label{FIG. 1}(Color online) (a) Powder XRD pattern with the refinement profile at room temperature. The inset: Crystal
		and magnetic structures. (b) Single-crystal XRD pattern and a photograph (inset) of a CrSb crystal. (c) Band structures with spin-orbit coupling (SOC). (d) Fermi surface calculated with SOC.}
\end{figure}

Single crystals of CrSb were grown using a chemical vapor transport method. Chromium and antimony powders were finely ground and sealed in a quartz tube under vacuum, with iodine (10 mg/cm$^3$) as the transport agent. The sealed tube was placed in a two-zone furnace, with the source end maintained at 950 $^{\circ}$C and the sink end at 850 $^{\circ}$C for two weeks. Shiny single crystals with typical dimensions of $2\times2\times0.05$ mm$^{3}$ were collected at the cooler end of the tube. Energy-dispersive x-ray spectroscopy (EDX) confirmed the stoichiometric ratio of Cr and Sb to be approximately 1: 1. The crystal structure was further characterized by powder x-ray diffraction (XRD, PANalytical) after grinding small portions of the crystals. Transport measurements, including electrical resistivity and Hall resistivity, were conducted using a Quantum Design physical property measurement system (PPMS-9 T), while magnetization measurements were performed with a magnetic property measurement system (MPMS-7 T). Longitudinal resistivity ($\rho_{xx}$) and transverse resistivity ($\rho_{yx}$) were measured using the standard four-terminal method, with the magnetic field polarity reversed to compensate for voltage contact misalignment. In parallel, numerical simulations based on Boltzmann transport theory and first-principles density functional theory (DFT) \cite{PhysRevB.99.035142} were performed to compare with the experimental results. DFT calculations were carried out using the Vienna ab initio simulation package (VASP)~\cite{PhysRevB.54.11169,PhysRevB.59.1758} with the Perdew-Burke-Ernzerhof (PBE) generalized gradient approximation (GGA) for the exchange-correlation functional~\cite{PhysRevLett.77.3865}. A plane-wave cutoff energy of 400 eV and an $11\times11\times9$ k-point mesh were used for the bulk calculations. The FS and magneto-transport properties were calculated using WannierTools \cite{WU2018405}, leveraging the Wannier tight-binding model (WTBM) \cite{PhysRevB.56.12847,PhysRevB.65.035109,RevModPhys.84.1419} constructed via Wannier90 \cite{MOSTOFI20142309}.


Figure 1(a) displays the polycrystalline x-ray diffraction (XRD) pattern along with the refinement profile and the crystal structure  of CrSb, which crystallizes a hexagonal structure of the NiAs-type with the space group $\emph{P}$$6_{3}/\emph{mmc}$ (No.~194). The Cr and Sb atoms are located at the Wyckoff positions 2a (0, 0, 0) and 2c (1/3, 2/3, 1/4), respectively. The lattice parameters $\textit{a}$ = $\textit{b}$ = 4.126(4) \r{A}, $\textit{c}$ = 5.469(3) \r{A} were obtained by fitting the XRD data using Rietveld refinement with a weighted profile factor R$_{wp}$ = 6.7$\%$ and a goodness of fit $\chi^{2}$ = 1.4. Figure 1(b) shows the single crystal XRD pattern, only (00$l$) peaks were observed, and the narrow half-angle width ($\Delta$$\theta$ $\sim$ 0.05°) indicates high-quality crystal. The calculated band structure and Fermi surfaces (FSs) are shown in Figs. 1(c) and 1(d), respectively, which is a clear metallic band and consistent with previous calculations and ARPES measurements \cite{li2024topological,lu2024observation,zeng2024observation}.

\begin{figure}
	\includegraphics[width= 8.6cm]{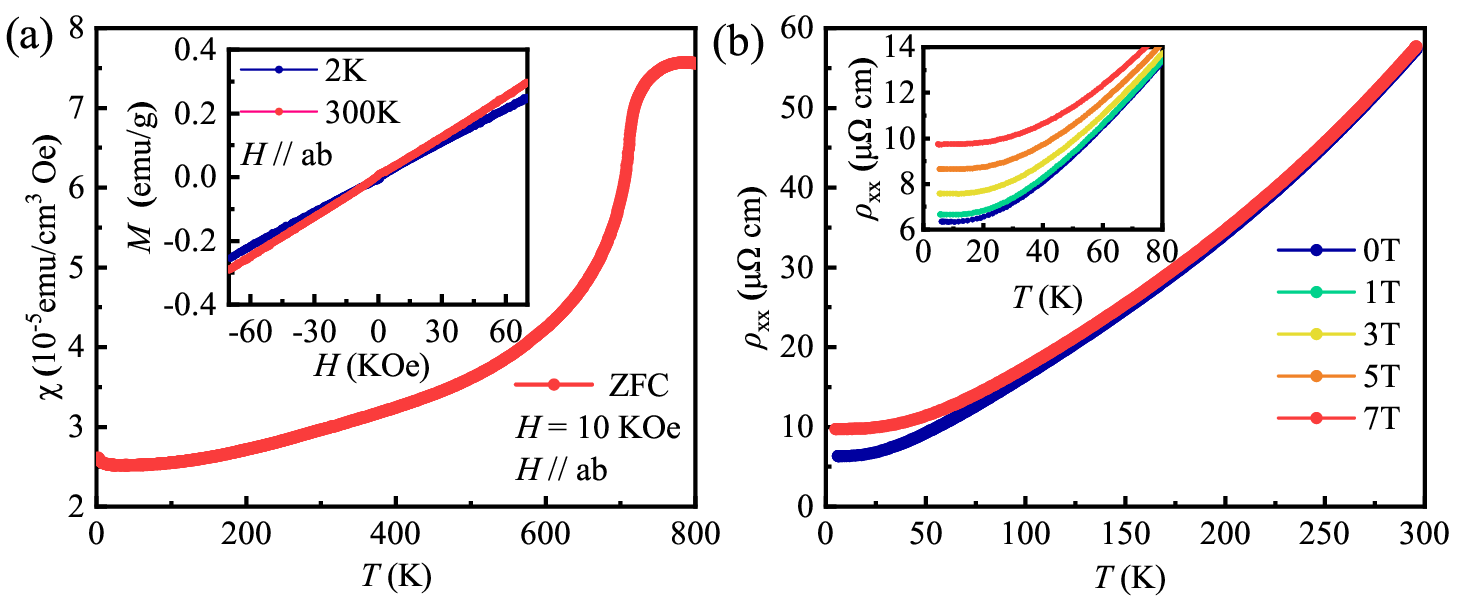}
	\caption{\label{FIG. 2}(Color online) (a) Temperature dependence of susceptibility  $\chi(T)$ measured at $\mu_{0}H$ = 1 T applied in the \textit{ab} plane with a zero-field-cooled (ZFC) process. The inset: Magnetization $M$ as a function of magnetic field measured at $T$ = 2 K and 300 K. (b) $\rho_{xx}(T)$ measured at various fields; the inset highlights the low-temperature data.}  
\end{figure}

Figure 2(a) shows the temperature dependence of the susceptibility, $\chi(T)$, measured at $\mu_{0}H$ = 1 T applied in the $ab$ plane with the zero-field-cooling process up to 800 K. With decreasing temperature, the $\chi$ remains almost a constant ($>$ 750 K), then drops sharply at about 712 K, subsequently becomes to declining gradually from 700 K to 20 K. The sharp drop at almost 712 K corresponds to an antiferromagnetic (AFM) transition ($T_{N}$ = 712 K, determined by the temperature of the peak in $d$$\chi(T)$/$d$T curve, not shown), consistent with previous reports \cite{PhysRev.85.365,PhysRev.129.2008}. The linear behavior in $M(H)$ curves measured at both 2 K and 300 K, as shown in the inset
of Fig. 2(a), indicates that the AFM state is stable below 700 K. Figure 2(b) presents the temperature dependence of longitudinal resistivity (in-plane) $\rho_{xx}(T)$ measured at  magnetic fields of 0 and 7 T. For the resistivity measured at zero field, as the temperature decreases, $\rho_{xx}(T)$ decreases monotonically from $\rho_{xx}$(300 K) = 57.5 $\mu$$\Omega$~cm to $\rho_{xx}$(5 K) = 6.43 $\mu$$\Omega$~cm, resulting in a residual resistivity ratio ($RRR$ = 8.94). We also measured the $\rho_{xx}(T)$ at various magnetic fields, as shown in the inset of Fig. 2(b), the  enhancement of $\rho_{xx}$ at lower temperatures indicates that the positive magnetoresistance also emerges in this AM crystal, reminiscent of extremely large magnetoresistance (XMR) observed in many topological nontrivial/trivial semimetals \cite{du2016large,PhysRevB.97.245101,PhysRevX.5.031023,PhysRevLett.111.056601,PhysRevB.85.035135,PhysRevB.94.235154,ali2014large}.

\begin{figure}
	\includegraphics[width= 8.6cm]{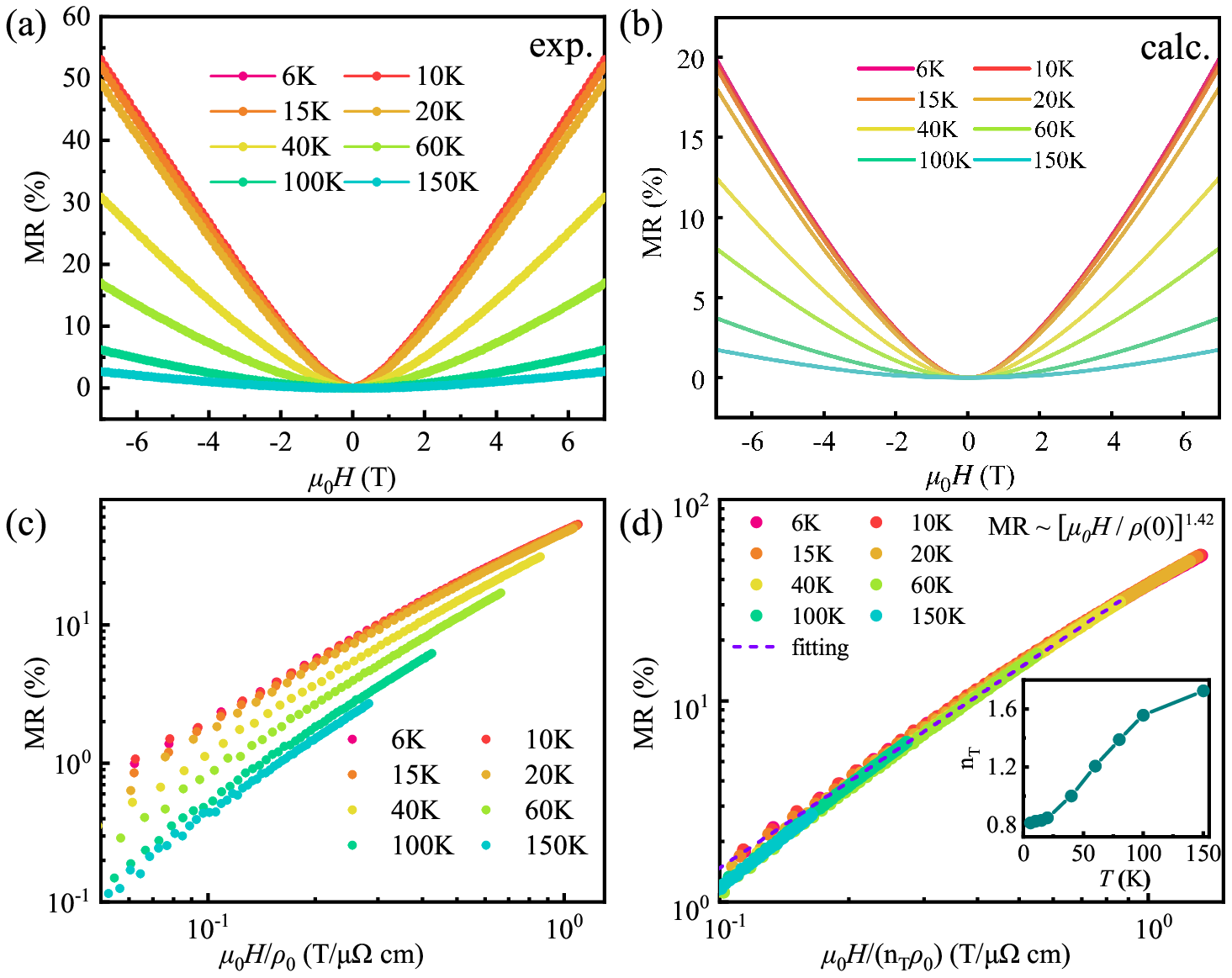}
	\caption{\label{FIG. 3} (a) The field dependence of MR measured at various temperatures, MR($H$).
     (b) The calculated MR($H$) by numerical simulations for different temperatures. (c) Kohler’s rule plots of the MR at various temperatures. (d) Extended Kohler’s rule plots of the MR, the inset presents the temperature dependence of \textit{n}$_{T}$.}
\end{figure}

Next, we focus on the MR scaling behavior, as done for many topological nontrivial/trivial semimetals \cite{PhysRevB.102.165133,PhysRevB.102.115145,PhysRevB.92.180402}. Figure 3(a) presents the MR as a function of magnetic field $\mu_{0}H$: MR($H$), measured at various temperatures, where MR is calculated by the standard definition MR = $\triangle\rho$/$\rho$(0) = [$\rho(H)$ - $\rho(0)$]/${\rho(0)}\times100\%$, $\rho(H)$ and $\rho(0)$ being the resistivity measured at $H$ and zero field, respectively. The MR reaches 52.6\% at 6 K and 7 T and does not saturate up to the highest magnetic field in our experiments, larger than that reported in Ref \cite{bai2024non}, similar to those discovered in many semimetals \cite{du2016large,PhysRevB.97.245101,PhysRevB.102.165133,PhysRevB.102.115145}. Figure 3(b) displays the numerical calculated MR($H$) results based on Boltzmann transport theory and the band structures without considering AFM order. It is clear that the calculated MR($H$) almost reproduces the main characteristics of the experimental MR($H$) curves at various temperatures, indicating that the magneto-transport properties originate from the intrinsic electronic structure and are dominated by the Lorentz force, even in this AFM material. At first, we try to analyze the MR data at various temperatures using a simple Kohler's rule: MR = $\alpha$($H/\rho_{0}$)$^{m}$, as done for many semimetals~\cite{PhysRevB.102.165133,PhysRevB.102.115145,PhysRevB.92.180402}. As shown in Fig. 3(c), no collapsing onto a single line for all the MR data demonstrates that the simple Kohler's rule cannot describe the MR data of CrSb crystal. Then, we reanalyze the MR data by using the extended Kohler's rule \cite{PhysRevX.11.041029,PhysRevB.110.205132}: MR = $\alpha(H/n_{T}\rho_{0})^{m}$, where $n_{T}$ is a temperature-dependent constant, related to the carrier density. We set $n_{T}$ = 1 at 40 K, and let all the MR data at various temperatures collapse onto the same line with that of 40 K by adjusting $n_{T}$ values. The obtained corresponding $n_{T}$ value for each temperature is shown in the inset of Fig. 3(d), ranging from $n_{T}$ = 0.8 at $T$ = 6 K to 1.72 at $T$ = 150 K. This analysis indicates that all the MR data can be well described by the extended Kohler's scale rule, MR = $\alpha(H/n_{T}\rho_{0})^{m}$, with $m$ = 1.42, consistent with the change of carrier density with temperature (see Fig. S1 in the Supplemental Material~\cite{SupplementalMaterial}).

\begin{figure}
	\includegraphics[width= 8.6cm]{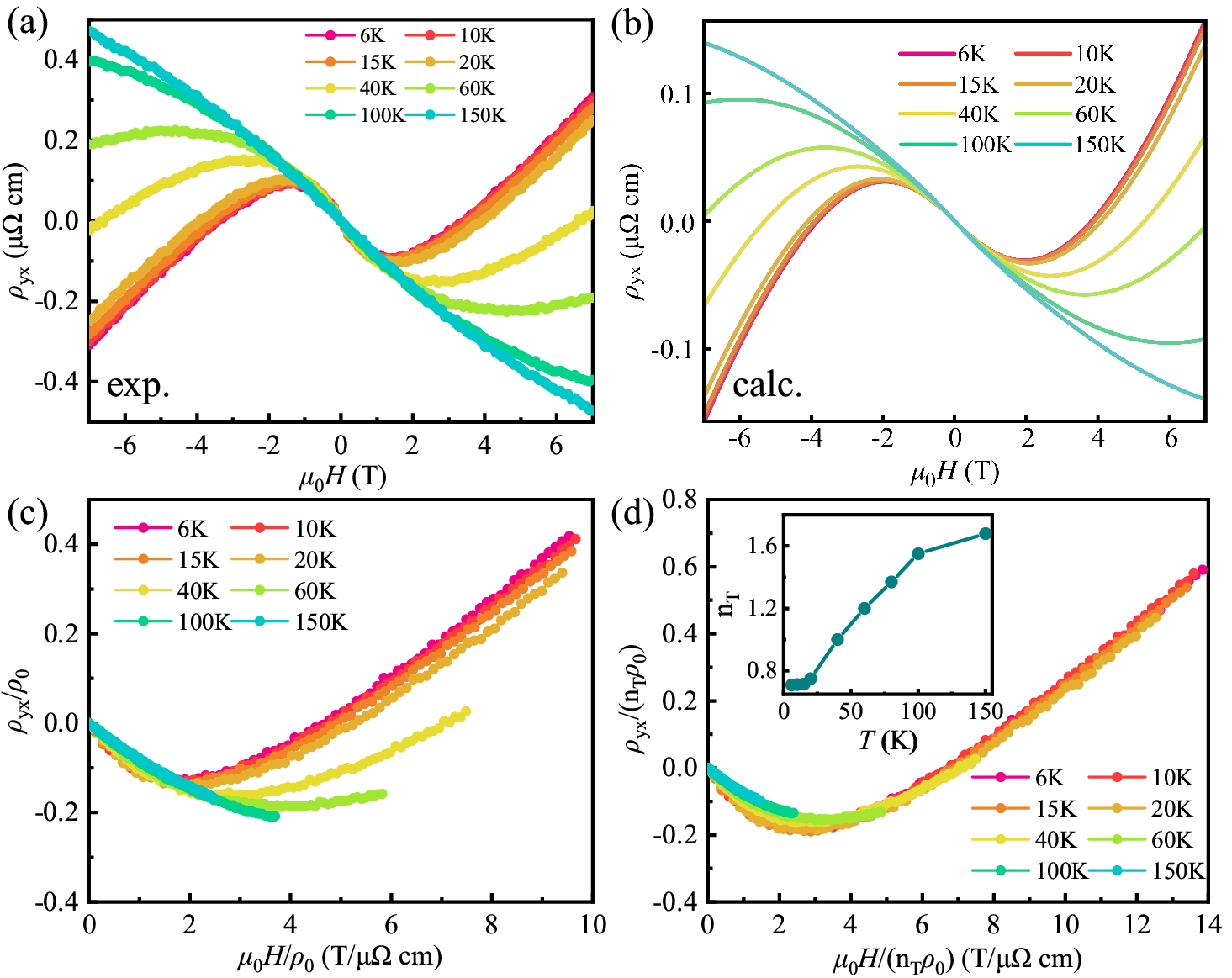}
	\caption{\label{FIG. 4}(a) Hall resistivity as a function of $H$, $\rho_{yx}(H)$, measured at various temperatures. (b) Numerical calculated $\rho_{yx}(H)$. (c) Scaling plots of $\rho_{yx}/\rho_{0}\sim H/\rho_{0}$ at various temperatures. (d) Scaling plots of $\rho_{yx}/(n_{T}\rho_{0}) \sim H/(n_{T}\rho_{0})$ at various temperatures.}
\end{figure}

Third, we discuss the scaling behavior in the Hall (transverse) resistivity. Figure 4(a) displays the Hall resistivity as a function of field measured at various temperatures, $\rho_{yx}(H)$, exhibiting distinctive behaviors at different temperatures. At 6 K, with increasing field from 0 to 7 T, $\rho_{yx}$ decreases at the beginning, reaches a minimum at 1.5 T, then increases, followed by a sign change from negative to positive at 3.5 T, similar behavior occurring in all the $\rho_{yx}$ below 60 K, and the nonlinear behavior of $\rho_{yx}$ remaining up to 150 K. Naturally, this nonlinear behavior of $\rho_{yx}(H)$ would be expected to be a ferromagnetic-like anomalous Hall effect (AHE), as observed in MnTe, another AM compound. However, as discussed by T.~Urata $et~al.$~\cite{urata2024high}, a spontaneous AHE can only be expected as the magnetic point group (MPG) is compatible with the pseudovector comprised of the antisymmetric Hall conductivity components~\cite{vsmejkal2020crystal}, $i.e.$, MPG allowing a ferromagnetic state. Although CrSb and MnTe share the same crystal structure, the MPG of CrSb (6$'$/$m'mm'$) is not compatible with ferromagnetism \cite{zhou2024crystal}. Another possibility is that the application of a magnetic field $\vec{H}$ may alter the MPG by rotating the Néel vector $\vec{L}$, resulting in an AHE. T.~Urata $et~al.$~\cite{urata2024high} have also excluded this possibility by using the crystal point group analysis~\cite{dzyaloshinskii1957}, and the above nonlinear behavior of the Hall resistivity cannot be attributed to the AHE. Then, we analyzed the $\rho_{yx}(H)$ data by using three-band (2 electron pockets and 1 hole pocket) model [see Fig. S1(a) in the Supplemental Material~\cite{SupplementalMaterial}], as discussed by Y.~Bai $et~al.$~\cite{bai2024non}. It was found that the three-band model can well describe all the $\rho_{yx}(H)$ data, as well as evidenced by the calculations based on the Boltzmann transport theory [see Fig. 4(b) and Fig. S1(b) in the Supplemental Material~\cite{SupplementalMaterial}].  

As discussed by us in Ref.~\cite{PhysRevB.110.205132}, the Hall resistivity may also follow a scaling law under certain conditions, especially when all charge carriers share the same temperature-dependent relaxation time, expressed as:
 \begin{equation}
 	\frac{\rho_{yx}}{n\rho_{0}/m} = h(\frac{B}{n\rho_{0}/m})
 \end{equation}
where $n$ and $m$ are carrier density and effective mass, respectively, and $\rho_0$ is the zero-field longitudinal resistivity.  This scaling suggests that, similar to Kohler's rule for longitudinal MR, the Hall resistivity can be described by a universal function when appropriately normalized. This finding provides a deeper understanding of magneto-transport phenomena and offers a unified framework for analyzing both longitudinal and transverse resistivity behaviors in materials. As discussed above for Kohler's rule of MR, we plotted $\rho_{yx}$/$\rho_{0}$ as a function of $H/\rho_{0}$, as shown in Fig. 4(c). It is clear that all the $\rho_{yx}$/$\rho_{0}$ data measured at various temperatures do not collapse onto a single line. We chose a temperature-dependent constant $n_{T}$ for each temperature with setting $n_{T}$  = 1.0 at 40 K, and let all the $\rho_{yx}$/($n_{T}\rho_{0}$) data below 150 K collapse onto a single line, as shown in Fig. 4(d), indicating that the Hall resistivity can be described by the scaling law expressed in Eq. (1). It is interesting that the chosen $n_{T}$ values for different temperatures, as shown in the inset of Fig. 4(d), are very close to those in the extended Kohler's rule for longitudinal MR above, implying that both longitudinal and transverse resistivity can be described within a unified framework. 

The scaling behavior of Hall resistivity has been infrequently examined, likely due to its complexity relative to Kohler’s rule for longitudinal resistivity. In many metals, semimetals, and semiconductors, Hall resistivity typically exhibits a linear dependence on the magnetic field with single charge carriers, and its characteristics remain largely temperature-independent in both low and high magnetic field regimes, thereby reducing the necessity of scaling. In contrast, materials with multiple types of charge carriers demonstrate a more intricate relationship between Hall resistivity, magnetic field, and temperature, often resulting in nonlinear features that can be misinterpreted as the AHE or indicative of new physics.

In conclusion, we successfully grew the AM CrSb single crystals and confirmed its AFM transition with $T_{N}$ = 712 K by high-temperature susceptibility measurements. By combining the electronic structure and FS calculations, as well as the magneto-transport numerical simulations based on Boltzmann transport theory, we studied the scaling law of longitudinal MR and Hall resistivity. It was found that the scaling law can very well describe the experimental data for this CrSb AM semimetal, consistent very well with the numerical simulation results. The nonlinear Hall resistivity results from its multi-band electronic structure, rather than being an anomalous Hall effect. These results demonstrate that the magneto-transport properties originate from the intrinsic electronic structure and are dominated by the Lorentz force even in this AM material.
~\\
\begin{acknowledgments}
This research is supported by the National Key R$\&$D program of China under Grant No. 2022YFA1403202, 2023YFA1607400 and the National Natural Science Foundation of China (Grant No. 12074335, 52471020, 12274436, 12188101).
\end{acknowledgments}

\nocite{*}


\begin{thebibliography}{51}%
	\makeatletter
	\providecommand \@ifxundefined [1]{%
		\@ifx{#1\undefined}
	}%
	\providecommand \@ifnum [1]{%
		\ifnum #1\expandafter \@firstoftwo
		\else \expandafter \@secondoftwo
		\fi
	}%
	\providecommand \@ifx [1]{%
		\ifx #1\expandafter \@firstoftwo
		\else \expandafter \@secondoftwo
		\fi
	}%
	\providecommand \natexlab [1]{#1}%
	\providecommand \enquote  [1]{``#1''}%
	\providecommand \bibnamefont  [1]{#1}%
	\providecommand \bibfnamefont [1]{#1}%
	\providecommand \citenamefont [1]{#1}%
	\providecommand \href@noop [0]{\@secondoftwo}%
	\providecommand \href [0]{\begingroup \@sanitize@url \@href}%
	\providecommand \@href[1]{\@@startlink{#1}\@@href}%
	\providecommand \@@href[1]{\endgroup#1\@@endlink}%
	\providecommand \@sanitize@url [0]{\catcode `\\12\catcode `\$12\catcode
		`\&12\catcode `\#12\catcode `\^12\catcode `\_12\catcode `\%12\relax}%
	\providecommand \@@startlink[1]{}%
	\providecommand \@@endlink[0]{}%
	\providecommand \url  [0]{\begingroup\@sanitize@url \@url }%
	\providecommand \@url [1]{\endgroup\@href {#1}{\urlprefix }}%
	\providecommand \urlprefix  [0]{URL }%
	\providecommand \Eprint [0]{\href }%
	\providecommand \doibase [0]{https://doi.org/}%
	\providecommand \selectlanguage [0]{\@gobble}%
	\providecommand \bibinfo  [0]{\@secondoftwo}%
	\providecommand \bibfield  [0]{\@secondoftwo}%
	\providecommand \translation [1]{[#1]}%
	\providecommand \BibitemOpen [0]{}%
	\providecommand \bibitemStop [0]{}%
	\providecommand \bibitemNoStop [0]{.\EOS\space}%
	\providecommand \EOS [0]{\spacefactor3000\relax}%
	\providecommand \BibitemShut  [1]{\csname bibitem#1\endcsname}%
	\let\auto@bib@innerbib\@empty
	\bibitem [{\citenamefont {Mazin}(2022)}]{PhysRevX.12.040002}%
	\BibitemOpen
	\bibfield  {author} {\bibinfo {author} {\bibfnamefont {I.}~\bibnamefont
			{Mazin}} (\bibinfo {collaboration} {The PRX Editors}),\ }\href
	{https://doi.org/10.1103/PhysRevX.12.040002} {\bibfield  {journal} {\bibinfo
			{journal} {Phys. Rev. X}\ }\textbf {\bibinfo {volume} {12}},\ \bibinfo
		{pages} {040002} (\bibinfo {year} {2022})}\BibitemShut {NoStop}%
	\bibitem [{\citenamefont {\ifmmode~\check{S}\else \v{S}\fi{}mejkal}\ \emph
		{et~al.}(2022{\natexlab{a}})\citenamefont {\ifmmode~\check{S}\else
			\v{S}\fi{}mejkal}, \citenamefont {Sinova},\ and\ \citenamefont
		{Jungwirth}}]{PhysRevX.12.031042}%
	\BibitemOpen
	\bibfield  {author} {\bibinfo {author} {\bibfnamefont {L.}~\bibnamefont
			{\ifmmode~\check{S}\else \v{S}\fi{}mejkal}}, \bibinfo {author} {\bibfnamefont
			{J.}~\bibnamefont {Sinova}},\ and\ \bibinfo {author} {\bibfnamefont
			{T.}~\bibnamefont {Jungwirth}},\ }\href
	{https://doi.org/10.1103/PhysRevX.12.031042} {\bibfield  {journal} {\bibinfo
			{journal} {Phys. Rev. X}\ }\textbf {\bibinfo {volume} {12}},\ \bibinfo
		{pages} {031042} (\bibinfo {year} {2022}{\natexlab{a}})}\BibitemShut
	{NoStop}%
	\bibitem [{\citenamefont {\ifmmode~\check{S}\else \v{S}\fi{}mejkal}\ \emph
		{et~al.}(2022{\natexlab{b}})\citenamefont {\ifmmode~\check{S}\else
			\v{S}\fi{}mejkal}, \citenamefont {Sinova},\ and\ \citenamefont
		{Jungwirth}}]{PhysRevX.12.040501}%
	\BibitemOpen
	\bibfield  {author} {\bibinfo {author} {\bibfnamefont {L.}~\bibnamefont
			{\ifmmode~\check{S}\else \v{S}\fi{}mejkal}}, \bibinfo {author} {\bibfnamefont
			{J.}~\bibnamefont {Sinova}},\ and\ \bibinfo {author} {\bibfnamefont
			{T.}~\bibnamefont {Jungwirth}},\ }\href
	{https://doi.org/10.1103/PhysRevX.12.040501} {\bibfield  {journal} {\bibinfo
			{journal} {Phys. Rev. X}\ }\textbf {\bibinfo {volume} {12}},\ \bibinfo
		{pages} {040501} (\bibinfo {year} {2022}{\natexlab{b}})}\BibitemShut
	{NoStop}%
	\bibitem [{\citenamefont {Lee}\ \emph {et~al.}(2024)\citenamefont {Lee},
		\citenamefont {Lee}, \citenamefont {Jung}, \citenamefont {Jung},
		\citenamefont {Kim}, \citenamefont {Lee}, \citenamefont {Seok}, \citenamefont
		{Kim}, \citenamefont {Park}, \citenamefont {\ifmmode~\check{S}\else
			\v{S}\fi{}mejkal}, \citenamefont {Kang},\ and\ \citenamefont
		{Kim}}]{PhysRevLett.132.036702}%
	\BibitemOpen
	\bibfield  {author} {\bibinfo {author} {\bibfnamefont {S.}~\bibnamefont
			{Lee}}, \bibinfo {author} {\bibfnamefont {S.}~\bibnamefont {Lee}}, \bibinfo
		{author} {\bibfnamefont {S.}~\bibnamefont {Jung}}, \bibinfo {author}
		{\bibfnamefont {J.}~\bibnamefont {Jung}}, \bibinfo {author} {\bibfnamefont
			{D.}~\bibnamefont {Kim}}, \bibinfo {author} {\bibfnamefont {Y.}~\bibnamefont
			{Lee}}, \bibinfo {author} {\bibfnamefont {B.}~\bibnamefont {Seok}}, \bibinfo
		{author} {\bibfnamefont {J.}~\bibnamefont {Kim}}, \bibinfo {author}
		{\bibfnamefont {B.~G.}\ \bibnamefont {Park}}, \bibinfo {author}
		{\bibfnamefont {L.}~\bibnamefont {\ifmmode~\check{S}\else \v{S}\fi{}mejkal}},
		\bibinfo {author} {\bibfnamefont {C.-J.}\ \bibnamefont {Kang}},\ and\
		\bibinfo {author} {\bibfnamefont {C.}~\bibnamefont {Kim}},\ }\href
	{https://doi.org/10.1103/PhysRevLett.132.036702} {\bibfield  {journal}
		{\bibinfo  {journal} {Phys. Rev. Lett.}\ }\textbf {\bibinfo {volume} {132}},\
		\bibinfo {pages} {036702} (\bibinfo {year} {2024})}\BibitemShut {NoStop}%
	\bibitem [{\citenamefont {Osumi}\ \emph {et~al.}(2024)\citenamefont {Osumi},
		\citenamefont {Souma}, \citenamefont {Aoyama}, \citenamefont {Yamauchi},
		\citenamefont {Honma}, \citenamefont {Nakayama}, \citenamefont {Takahashi},
		\citenamefont {Ohgushi},\ and\ \citenamefont {Sato}}]{PhysRevB.109.115102}%
	\BibitemOpen
	\bibfield  {author} {\bibinfo {author} {\bibfnamefont {T.}~\bibnamefont
			{Osumi}}, \bibinfo {author} {\bibfnamefont {S.}~\bibnamefont {Souma}},
		\bibinfo {author} {\bibfnamefont {T.}~\bibnamefont {Aoyama}}, \bibinfo
		{author} {\bibfnamefont {K.}~\bibnamefont {Yamauchi}}, \bibinfo {author}
		{\bibfnamefont {A.}~\bibnamefont {Honma}}, \bibinfo {author} {\bibfnamefont
			{K.}~\bibnamefont {Nakayama}}, \bibinfo {author} {\bibfnamefont
			{T.}~\bibnamefont {Takahashi}}, \bibinfo {author} {\bibfnamefont
			{K.}~\bibnamefont {Ohgushi}},\ and\ \bibinfo {author} {\bibfnamefont
			{T.}~\bibnamefont {Sato}},\ }\href
	{https://doi.org/10.1103/PhysRevB.109.115102} {\bibfield  {journal} {\bibinfo
			{journal} {Phys. Rev. B}\ }\textbf {\bibinfo {volume} {109}},\ \bibinfo
		{pages} {115102} (\bibinfo {year} {2024})}\BibitemShut {NoStop}%
	\bibitem [{\citenamefont {Krempask{\`y}}\ \emph {et~al.}(2024)\citenamefont
		{Krempask{\`y}}, \citenamefont {{\v{S}}mejkal}, \citenamefont {D’souza},
		\citenamefont {Hajlaoui}, \citenamefont {Springholz}, \citenamefont
		{Uhl{\'\i}{\v{r}}ov{\'a}}, \citenamefont {Alarab}, \citenamefont
		{Constantinou}, \citenamefont {Strocov}, \citenamefont {Usanov} \emph
		{et~al.}}]{krempasky2024altermagnetic}%
	\BibitemOpen
	\bibfield  {author} {\bibinfo {author} {\bibfnamefont {J.}~\bibnamefont
			{Krempask{\`y}}}, \bibinfo {author} {\bibfnamefont {L.}~\bibnamefont
			{{\v{S}}mejkal}}, \bibinfo {author} {\bibfnamefont {S.}~\bibnamefont
			{D’souza}}, \bibinfo {author} {\bibfnamefont {M.}~\bibnamefont {Hajlaoui}},
		\bibinfo {author} {\bibfnamefont {G.}~\bibnamefont {Springholz}}, \bibinfo
		{author} {\bibfnamefont {K.}~\bibnamefont {Uhl{\'\i}{\v{r}}ov{\'a}}},
		\bibinfo {author} {\bibfnamefont {F.}~\bibnamefont {Alarab}}, \bibinfo
		{author} {\bibfnamefont {P.}~\bibnamefont {Constantinou}}, \bibinfo {author}
		{\bibfnamefont {V.}~\bibnamefont {Strocov}}, \bibinfo {author} {\bibfnamefont
			{D.}~\bibnamefont {Usanov}}, \emph {et~al.},\ }\href
	{https://doi.org/https://doi.org/10.1038/s41586-023-06907-7} {\bibfield
		{journal} {\bibinfo  {journal} {Nature}\ }\textbf {\bibinfo {volume} {626}},\
		\bibinfo {pages} {517} (\bibinfo {year} {2024})}\BibitemShut {NoStop}%
	\bibitem [{\citenamefont {Reimers}\ \emph {et~al.}(2024)\citenamefont
		{Reimers}, \citenamefont {Odenbreit}, \citenamefont {{\v{S}}mejkal},
		\citenamefont {Strocov}, \citenamefont {Constantinou}, \citenamefont
		{Hellenes}, \citenamefont {Jaeschke~Ubiergo}, \citenamefont {Campos},
		\citenamefont {Bharadwaj}, \citenamefont {Chakraborty} \emph
		{et~al.}}]{reimers2024direct}%
	\BibitemOpen
	\bibfield  {author} {\bibinfo {author} {\bibfnamefont {S.}~\bibnamefont
			{Reimers}}, \bibinfo {author} {\bibfnamefont {L.}~\bibnamefont {Odenbreit}},
		\bibinfo {author} {\bibfnamefont {L.}~\bibnamefont {{\v{S}}mejkal}}, \bibinfo
		{author} {\bibfnamefont {V.~N.}\ \bibnamefont {Strocov}}, \bibinfo {author}
		{\bibfnamefont {P.}~\bibnamefont {Constantinou}}, \bibinfo {author}
		{\bibfnamefont {A.~B.}\ \bibnamefont {Hellenes}}, \bibinfo {author}
		{\bibfnamefont {R.}~\bibnamefont {Jaeschke~Ubiergo}}, \bibinfo {author}
		{\bibfnamefont {W.~H.}\ \bibnamefont {Campos}}, \bibinfo {author}
		{\bibfnamefont {V.~K.}\ \bibnamefont {Bharadwaj}}, \bibinfo {author}
		{\bibfnamefont {A.}~\bibnamefont {Chakraborty}}, \emph {et~al.},\ }\href
	{https://doi.org/10.1038/s41467-024-46476-5} {\bibfield  {journal} {\bibinfo
			{journal} {Nat. Commun.}\ }\textbf {\bibinfo {volume} {15}},\ \bibinfo
		{pages} {2116} (\bibinfo {year} {2024})}\BibitemShut {NoStop}%
	\bibitem [{\citenamefont {Fedchenko}\ \emph {et~al.}(2024)\citenamefont
		{Fedchenko}, \citenamefont {Min{\'a}r}, \citenamefont {Akashdeep},
		\citenamefont {D’Souza}, \citenamefont {Vasilyev}, \citenamefont {Tkach},
		\citenamefont {Odenbreit}, \citenamefont {Nguyen}, \citenamefont
		{Kutnyakhov}, \citenamefont {Wind} \emph {et~al.}}]{fedchenko2024}%
	\BibitemOpen
	\bibfield  {author} {\bibinfo {author} {\bibfnamefont {O.}~\bibnamefont
			{Fedchenko}}, \bibinfo {author} {\bibfnamefont {J.}~\bibnamefont
			{Min{\'a}r}}, \bibinfo {author} {\bibfnamefont {A.}~\bibnamefont
			{Akashdeep}}, \bibinfo {author} {\bibfnamefont {S.~W.}\ \bibnamefont
			{D’Souza}}, \bibinfo {author} {\bibfnamefont {D.}~\bibnamefont {Vasilyev}},
		\bibinfo {author} {\bibfnamefont {O.}~\bibnamefont {Tkach}}, \bibinfo
		{author} {\bibfnamefont {L.}~\bibnamefont {Odenbreit}}, \bibinfo {author}
		{\bibfnamefont {Q.}~\bibnamefont {Nguyen}}, \bibinfo {author} {\bibfnamefont
			{D.}~\bibnamefont {Kutnyakhov}}, \bibinfo {author} {\bibfnamefont
			{N.}~\bibnamefont {Wind}}, \emph {et~al.},\ }\href
	{https://doi.org/10.1126/sciadv.adj4883} {\bibfield  {journal} {\bibinfo
			{journal} {Sci. Adv.}\ }\textbf {\bibinfo {volume} {10}},\ \bibinfo {pages}
		{eadj4883} (\bibinfo {year} {2024})}\BibitemShut {NoStop}%
	\bibitem [{\citenamefont {Gonzalez~Betancourt}\ \emph
		{et~al.}(2023)\citenamefont {Gonzalez~Betancourt}, \citenamefont
		{Zub\'a\ifmmode~\check{c}\else \v{c}\fi{}}, \citenamefont
		{Gonzalez-Hernandez}, \citenamefont {Geishendorf}, \citenamefont {\ifmmode
			\check{S}\else \v{S}\fi{}ob\'a\ifmmode~\check{n}\else \v{n}\fi{}},
		\citenamefont {Springholz}, \citenamefont {Olejn\'{\i}k}, \citenamefont
		{\ifmmode~\check{S}\else \v{S}\fi{}mejkal}, \citenamefont {Sinova},
		\citenamefont {Jungwirth}, \citenamefont {Goennenwein}, \citenamefont
		{Thomas}, \citenamefont {Reichlov\'a}, \citenamefont {\ifmmode~\check{Z}\else
			\v{Z}\fi{}elezn\'y},\ and\ \citenamefont
		{Kriegner}}]{PhysRevLett.130.036702}%
	\BibitemOpen
	\bibfield  {author} {\bibinfo {author} {\bibfnamefont {R.~D.}\ \bibnamefont
			{Gonzalez~Betancourt}}, \bibinfo {author} {\bibfnamefont {J.}~\bibnamefont
			{Zub\'a\ifmmode~\check{c}\else \v{c}\fi{}}}, \bibinfo {author} {\bibfnamefont
			{R.}~\bibnamefont {Gonzalez-Hernandez}}, \bibinfo {author} {\bibfnamefont
			{K.}~\bibnamefont {Geishendorf}}, \bibinfo {author} {\bibfnamefont
			{Z.}~\bibnamefont {\ifmmode \check{S}\else
				\v{S}\fi{}ob\'a\ifmmode~\check{n}\else \v{n}\fi{}}}, \bibinfo {author}
		{\bibfnamefont {G.}~\bibnamefont {Springholz}}, \bibinfo {author}
		{\bibfnamefont {K.}~\bibnamefont {Olejn\'{\i}k}}, \bibinfo {author}
		{\bibfnamefont {L.}~\bibnamefont {\ifmmode~\check{S}\else \v{S}\fi{}mejkal}},
		\bibinfo {author} {\bibfnamefont {J.}~\bibnamefont {Sinova}}, \bibinfo
		{author} {\bibfnamefont {T.}~\bibnamefont {Jungwirth}}, \bibinfo {author}
		{\bibfnamefont {S.~T.~B.}\ \bibnamefont {Goennenwein}}, \bibinfo {author}
		{\bibfnamefont {A.}~\bibnamefont {Thomas}}, \bibinfo {author} {\bibfnamefont
			{H.}~\bibnamefont {Reichlov\'a}}, \bibinfo {author} {\bibfnamefont
			{J.}~\bibnamefont {\ifmmode~\check{Z}\else \v{Z}\fi{}elezn\'y}},\ and\
		\bibinfo {author} {\bibfnamefont {D.}~\bibnamefont {Kriegner}},\ }\href
	{https://doi.org/10.1103/PhysRevLett.130.036702} {\bibfield  {journal}
		{\bibinfo  {journal} {Phys. Rev. Lett.}\ }\textbf {\bibinfo {volume} {130}},\
		\bibinfo {pages} {036702} (\bibinfo {year} {2023})}\BibitemShut {NoStop}%
	\bibitem [{\citenamefont {Wang}\ \emph {et~al.}(2023)\citenamefont {Wang},
		\citenamefont {Tanaka}, \citenamefont {Sakai}, \citenamefont {Wang},
		\citenamefont {Deng}, \citenamefont {Lyu}, \citenamefont {Li}, \citenamefont
		{Tian}, \citenamefont {Shen}, \citenamefont {Ogawa} \emph
		{et~al.}}]{wang2023emergent}%
	\BibitemOpen
	\bibfield  {author} {\bibinfo {author} {\bibfnamefont {M.}~\bibnamefont
			{Wang}}, \bibinfo {author} {\bibfnamefont {K.}~\bibnamefont {Tanaka}},
		\bibinfo {author} {\bibfnamefont {S.}~\bibnamefont {Sakai}}, \bibinfo
		{author} {\bibfnamefont {Z.}~\bibnamefont {Wang}}, \bibinfo {author}
		{\bibfnamefont {K.}~\bibnamefont {Deng}}, \bibinfo {author} {\bibfnamefont
			{Y.}~\bibnamefont {Lyu}}, \bibinfo {author} {\bibfnamefont {C.}~\bibnamefont
			{Li}}, \bibinfo {author} {\bibfnamefont {D.}~\bibnamefont {Tian}}, \bibinfo
		{author} {\bibfnamefont {S.}~\bibnamefont {Shen}}, \bibinfo {author}
		{\bibfnamefont {N.}~\bibnamefont {Ogawa}}, \emph {et~al.},\ }\href
	{https://doi.org/10.1038/s41467-023-43962-0} {\bibfield  {journal} {\bibinfo
			{journal} {Nat. Commun.}\ }\textbf {\bibinfo {volume} {14}},\ \bibinfo
		{pages} {8240} (\bibinfo {year} {2023})}\BibitemShut {NoStop}%
	\bibitem [{\citenamefont {Feng}\ \emph {et~al.}(2022)\citenamefont {Feng},
		\citenamefont {Zhou}, \citenamefont {{\v{S}}mejkal}, \citenamefont {Wu},
		\citenamefont {Zhu}, \citenamefont {Guo}, \citenamefont
		{Gonz{\'a}lez-Hern{\'a}ndez}, \citenamefont {Wang}, \citenamefont {Yan},
		\citenamefont {Qin} \emph {et~al.}}]{feng2022anomalous}%
	\BibitemOpen
	\bibfield  {author} {\bibinfo {author} {\bibfnamefont {Z.}~\bibnamefont
			{Feng}}, \bibinfo {author} {\bibfnamefont {X.}~\bibnamefont {Zhou}}, \bibinfo
		{author} {\bibfnamefont {L.}~\bibnamefont {{\v{S}}mejkal}}, \bibinfo {author}
		{\bibfnamefont {L.}~\bibnamefont {Wu}}, \bibinfo {author} {\bibfnamefont
			{Z.}~\bibnamefont {Zhu}}, \bibinfo {author} {\bibfnamefont {H.}~\bibnamefont
			{Guo}}, \bibinfo {author} {\bibfnamefont {R.}~\bibnamefont
			{Gonz{\'a}lez-Hern{\'a}ndez}}, \bibinfo {author} {\bibfnamefont
			{X.}~\bibnamefont {Wang}}, \bibinfo {author} {\bibfnamefont {H.}~\bibnamefont
			{Yan}}, \bibinfo {author} {\bibfnamefont {P.}~\bibnamefont {Qin}}, \emph
		{et~al.},\ }\href {https://doi.org/10.1038/s41928-022-00866-z} {\bibfield
		{journal} {\bibinfo  {journal} {Nat. Electron.}\ }\textbf {\bibinfo {volume}
			{5}},\ \bibinfo {pages} {735} (\bibinfo {year} {2022})}\BibitemShut {NoStop}%
	\bibitem [{\citenamefont {Tschirner}\ \emph {et~al.}(2023)\citenamefont
		{Tschirner}, \citenamefont {Ke{\ss}ler}, \citenamefont {Gonzalez~Betancourt},
		\citenamefont {Kotte}, \citenamefont {Kriegner}, \citenamefont {B{\"u}chner},
		\citenamefont {Dufouleur}, \citenamefont {Kamp}, \citenamefont {Jovic},
		\citenamefont {Smejkal} \emph {et~al.}}]{tschirner2023saturation}%
	\BibitemOpen
	\bibfield  {author} {\bibinfo {author} {\bibfnamefont {T.}~\bibnamefont
			{Tschirner}}, \bibinfo {author} {\bibfnamefont {P.}~\bibnamefont
			{Ke{\ss}ler}}, \bibinfo {author} {\bibfnamefont {R.~D.}\ \bibnamefont
			{Gonzalez~Betancourt}}, \bibinfo {author} {\bibfnamefont {T.}~\bibnamefont
			{Kotte}}, \bibinfo {author} {\bibfnamefont {D.}~\bibnamefont {Kriegner}},
		\bibinfo {author} {\bibfnamefont {B.}~\bibnamefont {B{\"u}chner}}, \bibinfo
		{author} {\bibfnamefont {J.}~\bibnamefont {Dufouleur}}, \bibinfo {author}
		{\bibfnamefont {M.}~\bibnamefont {Kamp}}, \bibinfo {author} {\bibfnamefont
			{V.}~\bibnamefont {Jovic}}, \bibinfo {author} {\bibfnamefont
			{L.}~\bibnamefont {Smejkal}}, \emph {et~al.},\ }\href
	{https://doi.org/10.1063/5.0160335} {\bibfield  {journal} {\bibinfo
			{journal} {APL Mater.}\ }\textbf {\bibinfo {volume} {11}},\ \bibinfo {pages}
		{101103} (\bibinfo {year} {2023})}\BibitemShut {NoStop}%
	\bibitem [{\citenamefont {Bai}\ \emph {et~al.}(2022)\citenamefont {Bai},
		\citenamefont {Han}, \citenamefont {Feng}, \citenamefont {Zhou},
		\citenamefont {Su}, \citenamefont {Wang}, \citenamefont {Liao}, \citenamefont
		{Zhu}, \citenamefont {Chen}, \citenamefont {Pan}, \citenamefont {Fan},\ and\
		\citenamefont {Song}}]{PhysRevLett.128.197202}%
	\BibitemOpen
	\bibfield  {author} {\bibinfo {author} {\bibfnamefont {H.}~\bibnamefont
			{Bai}}, \bibinfo {author} {\bibfnamefont {L.}~\bibnamefont {Han}}, \bibinfo
		{author} {\bibfnamefont {X.~Y.}\ \bibnamefont {Feng}}, \bibinfo {author}
		{\bibfnamefont {Y.~J.}\ \bibnamefont {Zhou}}, \bibinfo {author}
		{\bibfnamefont {R.~X.}\ \bibnamefont {Su}}, \bibinfo {author} {\bibfnamefont
			{Q.}~\bibnamefont {Wang}}, \bibinfo {author} {\bibfnamefont {L.~Y.}\
			\bibnamefont {Liao}}, \bibinfo {author} {\bibfnamefont {W.~X.}\ \bibnamefont
			{Zhu}}, \bibinfo {author} {\bibfnamefont {X.~Z.}\ \bibnamefont {Chen}},
		\bibinfo {author} {\bibfnamefont {F.}~\bibnamefont {Pan}}, \bibinfo {author}
		{\bibfnamefont {X.~L.}\ \bibnamefont {Fan}},\ and\ \bibinfo {author}
		{\bibfnamefont {C.}~\bibnamefont {Song}},\ }\href
	{https://doi.org/10.1103/PhysRevLett.128.197202} {\bibfield  {journal}
		{\bibinfo  {journal} {Phys. Rev. Lett.}\ }\textbf {\bibinfo {volume} {128}},\
		\bibinfo {pages} {197202} (\bibinfo {year} {2022})}\BibitemShut {NoStop}%
	\bibitem [{\citenamefont {Bose}\ \emph {et~al.}(2022)\citenamefont {Bose},
		\citenamefont {Schreiber}, \citenamefont {Jain}, \citenamefont {Shao},
		\citenamefont {Nair}, \citenamefont {Sun}, \citenamefont {Zhang},
		\citenamefont {Muller}, \citenamefont {Tsymbal}, \citenamefont {Schlom} \emph
		{et~al.}}]{bose2022tilted}%
	\BibitemOpen
	\bibfield  {author} {\bibinfo {author} {\bibfnamefont {A.}~\bibnamefont
			{Bose}}, \bibinfo {author} {\bibfnamefont {N.~J.}\ \bibnamefont {Schreiber}},
		\bibinfo {author} {\bibfnamefont {R.}~\bibnamefont {Jain}}, \bibinfo {author}
		{\bibfnamefont {D.-F.}\ \bibnamefont {Shao}}, \bibinfo {author}
		{\bibfnamefont {H.~P.}\ \bibnamefont {Nair}}, \bibinfo {author}
		{\bibfnamefont {J.}~\bibnamefont {Sun}}, \bibinfo {author} {\bibfnamefont
			{X.~S.}\ \bibnamefont {Zhang}}, \bibinfo {author} {\bibfnamefont {D.~A.}\
			\bibnamefont {Muller}}, \bibinfo {author} {\bibfnamefont {E.~Y.}\
			\bibnamefont {Tsymbal}}, \bibinfo {author} {\bibfnamefont {D.~G.}\
			\bibnamefont {Schlom}}, \emph {et~al.},\ }\href
	{https://doi.org/10.1038/s41928-022-00744-8} {\bibfield  {journal} {\bibinfo
			{journal} {Nat. Electron.}\ }\textbf {\bibinfo {volume} {5}},\ \bibinfo
		{pages} {267} (\bibinfo {year} {2022})}\BibitemShut {NoStop}%
	\bibitem [{\citenamefont {Aoyama}\ and\ \citenamefont
		{Ohgushi}(2024)}]{PhysRevMaterials.8.L041402}%
	\BibitemOpen
	\bibfield  {author} {\bibinfo {author} {\bibfnamefont {T.}~\bibnamefont
			{Aoyama}}\ and\ \bibinfo {author} {\bibfnamefont {K.}~\bibnamefont
			{Ohgushi}},\ }\href {https://doi.org/10.1103/PhysRevMaterials.8.L041402}
	{\bibfield  {journal} {\bibinfo  {journal} {Phys. Rev. Mater.}\ }\textbf
		{\bibinfo {volume} {8}},\ \bibinfo {pages} {L041402} (\bibinfo {year}
		{2024})}\BibitemShut {NoStop}%
	\bibitem [{\citenamefont {Zhu}\ \emph {et~al.}(2024)\citenamefont {Zhu},
		\citenamefont {Chen}, \citenamefont {Liu}, \citenamefont {Liu}, \citenamefont
		{Liu}, \citenamefont {Zha}, \citenamefont {Qu}, \citenamefont {Hong},
		\citenamefont {Li}, \citenamefont {Jiang} \emph
		{et~al.}}]{zhu2024observation}%
	\BibitemOpen
	\bibfield  {author} {\bibinfo {author} {\bibfnamefont {Y.-P.}\ \bibnamefont
			{Zhu}}, \bibinfo {author} {\bibfnamefont {X.}~\bibnamefont {Chen}}, \bibinfo
		{author} {\bibfnamefont {X.-R.}\ \bibnamefont {Liu}}, \bibinfo {author}
		{\bibfnamefont {Y.}~\bibnamefont {Liu}}, \bibinfo {author} {\bibfnamefont
			{P.}~\bibnamefont {Liu}}, \bibinfo {author} {\bibfnamefont {H.}~\bibnamefont
			{Zha}}, \bibinfo {author} {\bibfnamefont {G.}~\bibnamefont {Qu}}, \bibinfo
		{author} {\bibfnamefont {C.}~\bibnamefont {Hong}}, \bibinfo {author}
		{\bibfnamefont {J.}~\bibnamefont {Li}}, \bibinfo {author} {\bibfnamefont
			{Z.}~\bibnamefont {Jiang}}, \emph {et~al.},\ }\href
	{https://doi.org/10.1038/s41586-024-07023-w} {\bibfield  {journal} {\bibinfo
			{journal} {Nature}\ }\textbf {\bibinfo {volume} {626}},\ \bibinfo {pages}
		{523} (\bibinfo {year} {2024})}\BibitemShut {NoStop}%
	\bibitem [{\citenamefont {Haraldsen}\ \emph {et~al.}(1946)\citenamefont
		{Haraldsen}, \citenamefont {Rosenqvist},\ and\ \citenamefont
		{Gr{\o}Nvold}}]{haraldsen1946system}%
	\BibitemOpen
	\bibfield  {author} {\bibinfo {author} {\bibfnamefont {H.}~\bibnamefont
			{Haraldsen}}, \bibinfo {author} {\bibfnamefont {T.}~\bibnamefont
			{Rosenqvist}},\ and\ \bibinfo {author} {\bibfnamefont {F.}~\bibnamefont
			{Gr{\o}Nvold}},\ }\href@noop {} {}\ (\bibinfo  {publisher} {Cammermeyer},\
	\bibinfo {year} {1946})\BibitemShut {NoStop}%
	\bibitem [{\citenamefont {Snow}(1952)}]{PhysRev.85.365}%
	\BibitemOpen
	\bibfield  {author} {\bibinfo {author} {\bibfnamefont {A.~I.}\ \bibnamefont
			{Snow}},\ }\href {https://doi.org/10.1103/PhysRev.85.365} {\bibfield
		{journal} {\bibinfo  {journal} {Phys. Rev.}\ }\textbf {\bibinfo {volume}
			{85}},\ \bibinfo {pages} {365} (\bibinfo {year} {1952})}\BibitemShut
	{NoStop}%
	\bibitem [{\citenamefont {Takei}\ \emph {et~al.}(1963)\citenamefont {Takei},
		\citenamefont {Cox},\ and\ \citenamefont {Shirane}}]{PhysRev.129.2008}%
	\BibitemOpen
	\bibfield  {author} {\bibinfo {author} {\bibfnamefont {W.~J.}\ \bibnamefont
			{Takei}}, \bibinfo {author} {\bibfnamefont {D.~E.}\ \bibnamefont {Cox}},\
		and\ \bibinfo {author} {\bibfnamefont {G.}~\bibnamefont {Shirane}},\ }\href
	{https://doi.org/10.1103/PhysRev.129.2008} {\bibfield  {journal} {\bibinfo
			{journal} {Phys. Rev.}\ }\textbf {\bibinfo {volume} {129}},\ \bibinfo {pages}
		{2008} (\bibinfo {year} {1963})}\BibitemShut {NoStop}%
	\bibitem [{\citenamefont {Guo}\ \emph {et~al.}(2023)\citenamefont {Guo},
		\citenamefont {Liu}, \citenamefont {Janson}, \citenamefont {Fulga},
		\citenamefont {van~den Brink},\ and\ \citenamefont {Facio}}]{guo2023spin}%
	\BibitemOpen
	\bibfield  {author} {\bibinfo {author} {\bibfnamefont {Y.}~\bibnamefont
			{Guo}}, \bibinfo {author} {\bibfnamefont {H.}~\bibnamefont {Liu}}, \bibinfo
		{author} {\bibfnamefont {O.}~\bibnamefont {Janson}}, \bibinfo {author}
		{\bibfnamefont {I.~C.}\ \bibnamefont {Fulga}}, \bibinfo {author}
		{\bibfnamefont {J.}~\bibnamefont {van~den Brink}},\ and\ \bibinfo {author}
		{\bibfnamefont {J.~I.}\ \bibnamefont {Facio}},\ }\href
	{https://doi.org/10.1016/j.mtphys.2023.100991} {\bibfield  {journal}
		{\bibinfo  {journal} {Materials Today Physics}\ }\textbf {\bibinfo {volume}
			{32}},\ \bibinfo {pages} {100991} (\bibinfo {year} {2023})}\BibitemShut
	{NoStop}%
	\bibitem [{\citenamefont {Ding}\ \emph {et~al.}(2024)\citenamefont {Ding},
		\citenamefont {Jiang}, \citenamefont {Chen}, \citenamefont {Tao},
		\citenamefont {Liu}, \citenamefont {Li}, \citenamefont {Liu}, \citenamefont
		{Sun}, \citenamefont {Cheng}, \citenamefont {Liu}, \citenamefont {Yang},
		\citenamefont {Zhang}, \citenamefont {Deng}, \citenamefont {Jing},
		\citenamefont {Huang}, \citenamefont {Shi}, \citenamefont {Ye}, \citenamefont
		{Qiao}, \citenamefont {Wang}, \citenamefont {Guo}, \citenamefont {Feng},\
		and\ \citenamefont {Shen}}]{PhysRevLett.133.206401}%
	\BibitemOpen
	\bibfield  {author} {\bibinfo {author} {\bibfnamefont {J.}~\bibnamefont
			{Ding}}, \bibinfo {author} {\bibfnamefont {Z.}~\bibnamefont {Jiang}},
		\bibinfo {author} {\bibfnamefont {X.}~\bibnamefont {Chen}}, \bibinfo {author}
		{\bibfnamefont {Z.}~\bibnamefont {Tao}}, \bibinfo {author} {\bibfnamefont
			{Z.}~\bibnamefont {Liu}}, \bibinfo {author} {\bibfnamefont {T.}~\bibnamefont
			{Li}}, \bibinfo {author} {\bibfnamefont {J.}~\bibnamefont {Liu}}, \bibinfo
		{author} {\bibfnamefont {J.}~\bibnamefont {Sun}}, \bibinfo {author}
		{\bibfnamefont {J.}~\bibnamefont {Cheng}}, \bibinfo {author} {\bibfnamefont
			{J.}~\bibnamefont {Liu}}, \bibinfo {author} {\bibfnamefont {Y.}~\bibnamefont
			{Yang}}, \bibinfo {author} {\bibfnamefont {R.}~\bibnamefont {Zhang}},
		\bibinfo {author} {\bibfnamefont {L.}~\bibnamefont {Deng}}, \bibinfo {author}
		{\bibfnamefont {W.}~\bibnamefont {Jing}}, \bibinfo {author} {\bibfnamefont
			{Y.}~\bibnamefont {Huang}}, \bibinfo {author} {\bibfnamefont
			{Y.}~\bibnamefont {Shi}}, \bibinfo {author} {\bibfnamefont {M.}~\bibnamefont
			{Ye}}, \bibinfo {author} {\bibfnamefont {S.}~\bibnamefont {Qiao}}, \bibinfo
		{author} {\bibfnamefont {Y.}~\bibnamefont {Wang}}, \bibinfo {author}
		{\bibfnamefont {Y.}~\bibnamefont {Guo}}, \bibinfo {author} {\bibfnamefont
			{D.}~\bibnamefont {Feng}},\ and\ \bibinfo {author} {\bibfnamefont
			{D.}~\bibnamefont {Shen}},\ }\href
	{https://doi.org/10.1103/PhysRevLett.133.206401} {\bibfield  {journal}
		{\bibinfo  {journal} {Phys. Rev. Lett.}\ }\textbf {\bibinfo {volume} {133}},\
		\bibinfo {pages} {206401} (\bibinfo {year} {2024})}\BibitemShut {NoStop}%
	\bibitem [{\citenamefont {Zhang}\ \emph {et~al.}(2019)\citenamefont {Zhang},
		\citenamefont {Wu}, \citenamefont {Liu},\ and\ \citenamefont
		{Yazyev}}]{PhysRevB.99.035142}%
	\BibitemOpen
	\bibfield  {author} {\bibinfo {author} {\bibfnamefont {S.}~\bibnamefont
			{Zhang}}, \bibinfo {author} {\bibfnamefont {Q.}~\bibnamefont {Wu}}, \bibinfo
		{author} {\bibfnamefont {Y.}~\bibnamefont {Liu}},\ and\ \bibinfo {author}
		{\bibfnamefont {O.~V.}\ \bibnamefont {Yazyev}},\ }\href
	{https://doi.org/10.1103/PhysRevB.99.035142} {\bibfield  {journal} {\bibinfo
			{journal} {Phys. Rev. B}\ }\textbf {\bibinfo {volume} {99}},\ \bibinfo
		{pages} {035142} (\bibinfo {year} {2019})}\BibitemShut {NoStop}%
	\bibitem [{\citenamefont {Kresse}\ and\ \citenamefont
		{Furthm\"uller}(1996)}]{PhysRevB.54.11169}%
	\BibitemOpen
	\bibfield  {author} {\bibinfo {author} {\bibfnamefont {G.}~\bibnamefont
			{Kresse}}\ and\ \bibinfo {author} {\bibfnamefont {J.}~\bibnamefont
			{Furthm\"uller}},\ }\href {https://doi.org/10.1103/PhysRevB.54.11169}
	{\bibfield  {journal} {\bibinfo  {journal} {Phys. Rev. B}\ }\textbf {\bibinfo
			{volume} {54}},\ \bibinfo {pages} {11169} (\bibinfo {year}
		{1996})}\BibitemShut {NoStop}%
	\bibitem [{\citenamefont {Kresse}\ and\ \citenamefont
		{Joubert}(1999)}]{PhysRevB.59.1758}%
	\BibitemOpen
	\bibfield  {author} {\bibinfo {author} {\bibfnamefont {G.}~\bibnamefont
			{Kresse}}\ and\ \bibinfo {author} {\bibfnamefont {D.}~\bibnamefont
			{Joubert}},\ }\href {https://doi.org/10.1103/PhysRevB.59.1758} {\bibfield
		{journal} {\bibinfo  {journal} {Phys. Rev. B}\ }\textbf {\bibinfo {volume}
			{59}},\ \bibinfo {pages} {1758} (\bibinfo {year} {1999})}\BibitemShut
	{NoStop}%
	\bibitem [{\citenamefont {Perdew}\ \emph {et~al.}(1996)\citenamefont {Perdew},
		\citenamefont {Burke},\ and\ \citenamefont
		{Ernzerhof}}]{PhysRevLett.77.3865}%
	\BibitemOpen
	\bibfield  {author} {\bibinfo {author} {\bibfnamefont {J.~P.}\ \bibnamefont
			{Perdew}}, \bibinfo {author} {\bibfnamefont {K.}~\bibnamefont {Burke}},\ and\
		\bibinfo {author} {\bibfnamefont {M.}~\bibnamefont {Ernzerhof}},\ }\href
	{https://doi.org/10.1103/PhysRevLett.77.3865} {\bibfield  {journal} {\bibinfo
			{journal} {Phys. Rev. Lett.}\ }\textbf {\bibinfo {volume} {77}},\ \bibinfo
		{pages} {3865} (\bibinfo {year} {1996})}\BibitemShut {NoStop}%
	\bibitem [{\citenamefont {Wu}\ \emph {et~al.}(2018)\citenamefont {Wu},
		\citenamefont {Zhang}, \citenamefont {Song}, \citenamefont {Troyer},\ and\
		\citenamefont {Soluyanov}}]{WU2018405}%
	\BibitemOpen
	\bibfield  {author} {\bibinfo {author} {\bibfnamefont {Q.}~\bibnamefont
			{Wu}}, \bibinfo {author} {\bibfnamefont {S.}~\bibnamefont {Zhang}}, \bibinfo
		{author} {\bibfnamefont {H.-F.}\ \bibnamefont {Song}}, \bibinfo {author}
		{\bibfnamefont {M.}~\bibnamefont {Troyer}},\ and\ \bibinfo {author}
		{\bibfnamefont {A.~A.}\ \bibnamefont {Soluyanov}},\ }\href
	{https://doi.org/https://doi.org/10.1016/j.cpc.2017.09.033} {\bibfield
		{journal} {\bibinfo  {journal} {Comput. Phys. Commun.}\ }\textbf {\bibinfo
			{volume} {224}},\ \bibinfo {pages} {405} (\bibinfo {year}
		{2018})}\BibitemShut {NoStop}%
	\bibitem [{\citenamefont {Marzari}\ and\ \citenamefont
		{Vanderbilt}(1997)}]{PhysRevB.56.12847}%
	\BibitemOpen
	\bibfield  {author} {\bibinfo {author} {\bibfnamefont {N.}~\bibnamefont
			{Marzari}}\ and\ \bibinfo {author} {\bibfnamefont {D.}~\bibnamefont
			{Vanderbilt}},\ }\href {https://doi.org/10.1103/PhysRevB.56.12847} {\bibfield
		{journal} {\bibinfo  {journal} {Phys. Rev. B}\ }\textbf {\bibinfo {volume}
			{56}},\ \bibinfo {pages} {12847} (\bibinfo {year} {1997})}\BibitemShut
	{NoStop}%
	\bibitem [{\citenamefont {Souza}\ \emph {et~al.}(2001)\citenamefont {Souza},
		\citenamefont {Marzari},\ and\ \citenamefont
		{Vanderbilt}}]{PhysRevB.65.035109}%
	\BibitemOpen
	\bibfield  {author} {\bibinfo {author} {\bibfnamefont {I.}~\bibnamefont
			{Souza}}, \bibinfo {author} {\bibfnamefont {N.}~\bibnamefont {Marzari}},\
		and\ \bibinfo {author} {\bibfnamefont {D.}~\bibnamefont {Vanderbilt}},\
	}\href {https://doi.org/10.1103/PhysRevB.65.035109} {\bibfield  {journal}
		{\bibinfo  {journal} {Phys. Rev. B}\ }\textbf {\bibinfo {volume} {65}},\
		\bibinfo {pages} {035109} (\bibinfo {year} {2001})}\BibitemShut {NoStop}%
	\bibitem [{\citenamefont {Marzari}\ \emph {et~al.}(2012)\citenamefont
		{Marzari}, \citenamefont {Mostofi}, \citenamefont {Yates}, \citenamefont
		{Souza},\ and\ \citenamefont {Vanderbilt}}]{RevModPhys.84.1419}%
	\BibitemOpen
	\bibfield  {author} {\bibinfo {author} {\bibfnamefont {N.}~\bibnamefont
			{Marzari}}, \bibinfo {author} {\bibfnamefont {A.~A.}\ \bibnamefont
			{Mostofi}}, \bibinfo {author} {\bibfnamefont {J.~R.}\ \bibnamefont {Yates}},
		\bibinfo {author} {\bibfnamefont {I.}~\bibnamefont {Souza}},\ and\ \bibinfo
		{author} {\bibfnamefont {D.}~\bibnamefont {Vanderbilt}},\ }\href
	{https://doi.org/10.1103/RevModPhys.84.1419} {\bibfield  {journal} {\bibinfo
			{journal} {Rev. Mod. Phys.}\ }\textbf {\bibinfo {volume} {84}},\ \bibinfo
		{pages} {1419} (\bibinfo {year} {2012})}\BibitemShut {NoStop}%
	\bibitem [{\citenamefont {Mostofi}\ \emph {et~al.}(2014)\citenamefont
		{Mostofi}, \citenamefont {Yates}, \citenamefont {Pizzi}, \citenamefont {Lee},
		\citenamefont {Souza}, \citenamefont {Vanderbilt},\ and\ \citenamefont
		{Marzari}}]{MOSTOFI20142309}%
	\BibitemOpen
	\bibfield  {author} {\bibinfo {author} {\bibfnamefont {A.~A.}\ \bibnamefont
			{Mostofi}}, \bibinfo {author} {\bibfnamefont {J.~R.}\ \bibnamefont {Yates}},
		\bibinfo {author} {\bibfnamefont {G.}~\bibnamefont {Pizzi}}, \bibinfo
		{author} {\bibfnamefont {Y.-S.}\ \bibnamefont {Lee}}, \bibinfo {author}
		{\bibfnamefont {I.}~\bibnamefont {Souza}}, \bibinfo {author} {\bibfnamefont
			{D.}~\bibnamefont {Vanderbilt}},\ and\ \bibinfo {author} {\bibfnamefont
			{N.}~\bibnamefont {Marzari}},\ }\href
	{https://doi.org/https://doi.org/10.1016/j.cpc.2014.05.003} {\bibfield
		{journal} {\bibinfo  {journal} {Comput. Phys. Commun.}\ }\textbf {\bibinfo
			{volume} {185}},\ \bibinfo {pages} {2309} (\bibinfo {year}
		{2014})}\BibitemShut {NoStop}%
	\bibitem [{\citenamefont {Li}\ \emph {et~al.}(2024)\citenamefont {Li},
		\citenamefont {Hu}, \citenamefont {Li}, \citenamefont {Wang}, \citenamefont
		{Chen}, \citenamefont {Thiagarajan}, \citenamefont {Leandersson},
		\citenamefont {Polley}, \citenamefont {Kim}, \citenamefont {Liu},
		\citenamefont {Fulga}, \citenamefont {Vergniory}, \citenamefont {Janson},
		\citenamefont {Tjernberg},\ and\ \citenamefont {van~den
			Brink}}]{li2024topological}%
	\BibitemOpen
	\bibfield  {author} {\bibinfo {author} {\bibfnamefont {C.}~\bibnamefont
			{Li}}, \bibinfo {author} {\bibfnamefont {M.}~\bibnamefont {Hu}}, \bibinfo
		{author} {\bibfnamefont {Z.}~\bibnamefont {Li}}, \bibinfo {author}
		{\bibfnamefont {Y.}~\bibnamefont {Wang}}, \bibinfo {author} {\bibfnamefont
			{W.}~\bibnamefont {Chen}}, \bibinfo {author} {\bibfnamefont {B.}~\bibnamefont
			{Thiagarajan}}, \bibinfo {author} {\bibfnamefont {M.}~\bibnamefont
			{Leandersson}}, \bibinfo {author} {\bibfnamefont {C.}~\bibnamefont {Polley}},
		\bibinfo {author} {\bibfnamefont {T.}~\bibnamefont {Kim}}, \bibinfo {author}
		{\bibfnamefont {H.}~\bibnamefont {Liu}}, \bibinfo {author} {\bibfnamefont
			{C.}~\bibnamefont {Fulga}}, \bibinfo {author} {\bibfnamefont {M.~G.}\
			\bibnamefont {Vergniory}}, \bibinfo {author} {\bibfnamefont {O.}~\bibnamefont
			{Janson}}, \bibinfo {author} {\bibfnamefont {O.}~\bibnamefont {Tjernberg}},\
		and\ \bibinfo {author} {\bibfnamefont {J.}~\bibnamefont {van~den Brink}},\
	}\href {https://arxiv.org/abs/2405.14777} {} (\bibinfo {year} {2024}),\
	\Eprint {https://arxiv.org/abs/2405.14777} {arXiv:2405.14777
		[cond-mat.mtrl-sci]} \BibitemShut {NoStop}%
	\bibitem [{\citenamefont {Lu}\ \emph {et~al.}(2024)\citenamefont {Lu},
		\citenamefont {Feng}, \citenamefont {Wang}, \citenamefont {Chen},
		\citenamefont {Lin}, \citenamefont {Liang}, \citenamefont {Liu},
		\citenamefont {Feng}, \citenamefont {Yamagami}, \citenamefont {Liu},
		\citenamefont {Felser}, \citenamefont {Wu},\ and\ \citenamefont
		{Ma}}]{lu2024observation}%
	\BibitemOpen
	\bibfield  {author} {\bibinfo {author} {\bibfnamefont {W.}~\bibnamefont
			{Lu}}, \bibinfo {author} {\bibfnamefont {S.}~\bibnamefont {Feng}}, \bibinfo
		{author} {\bibfnamefont {Y.}~\bibnamefont {Wang}}, \bibinfo {author}
		{\bibfnamefont {D.}~\bibnamefont {Chen}}, \bibinfo {author} {\bibfnamefont
			{Z.}~\bibnamefont {Lin}}, \bibinfo {author} {\bibfnamefont {X.}~\bibnamefont
			{Liang}}, \bibinfo {author} {\bibfnamefont {S.}~\bibnamefont {Liu}}, \bibinfo
		{author} {\bibfnamefont {W.}~\bibnamefont {Feng}}, \bibinfo {author}
		{\bibfnamefont {K.}~\bibnamefont {Yamagami}}, \bibinfo {author}
		{\bibfnamefont {J.}~\bibnamefont {Liu}}, \bibinfo {author} {\bibfnamefont
			{C.}~\bibnamefont {Felser}}, \bibinfo {author} {\bibfnamefont
			{Q.}~\bibnamefont {Wu}},\ and\ \bibinfo {author} {\bibfnamefont
			{J.}~\bibnamefont {Ma}},\ }\href {https://arxiv.org/abs/2407.13497} {}
	(\bibinfo {year} {2024}),\ \Eprint {https://arxiv.org/abs/2407.13497}
	{arXiv:2407.13497 [cond-mat.mtrl-sci]} \BibitemShut {NoStop}%
	\bibitem [{\citenamefont {Zeng}\ \emph {et~al.}(2024)\citenamefont {Zeng},
		\citenamefont {Zhu}, \citenamefont {Zhu}, \citenamefont {Liu}, \citenamefont
		{Ma}, \citenamefont {Hao}, \citenamefont {Liu}, \citenamefont {Qu},
		\citenamefont {Yang}, \citenamefont {Jiang} \emph
		{et~al.}}]{zeng2024observation}%
	\BibitemOpen
	\bibfield  {author} {\bibinfo {author} {\bibfnamefont {M.}~\bibnamefont
			{Zeng}}, \bibinfo {author} {\bibfnamefont {M.-Y.}\ \bibnamefont {Zhu}},
		\bibinfo {author} {\bibfnamefont {Y.-P.}\ \bibnamefont {Zhu}}, \bibinfo
		{author} {\bibfnamefont {X.-R.}\ \bibnamefont {Liu}}, \bibinfo {author}
		{\bibfnamefont {X.-M.}\ \bibnamefont {Ma}}, \bibinfo {author} {\bibfnamefont
			{Y.-J.}\ \bibnamefont {Hao}}, \bibinfo {author} {\bibfnamefont
			{P.}~\bibnamefont {Liu}}, \bibinfo {author} {\bibfnamefont {G.}~\bibnamefont
			{Qu}}, \bibinfo {author} {\bibfnamefont {Y.}~\bibnamefont {Yang}}, \bibinfo
		{author} {\bibfnamefont {Z.}~\bibnamefont {Jiang}}, \emph {et~al.},\ }\href
	{https://doi.org/10.1002/advs.202406529} {\bibfield  {journal} {\bibinfo
			{journal} {Adv. Sci.}\ }\textbf {\bibinfo {volume} {11}},\ \bibinfo {pages}
		{2406529} (\bibinfo {year} {2024})}\BibitemShut {NoStop}%
	\bibitem [{\citenamefont {Du}\ \emph {et~al.}(2016)\citenamefont {Du},
		\citenamefont {Wang}, \citenamefont {Chen}, \citenamefont {Mao},
		\citenamefont {Khan}, \citenamefont {Xu}, \citenamefont {Zhou}, \citenamefont
		{Zhang}, \citenamefont {Yang}, \citenamefont {Chen} \emph
		{et~al.}}]{du2016large}%
	\BibitemOpen
	\bibfield  {author} {\bibinfo {author} {\bibfnamefont {J.}~\bibnamefont
			{Du}}, \bibinfo {author} {\bibfnamefont {H.}~\bibnamefont {Wang}}, \bibinfo
		{author} {\bibfnamefont {Q.}~\bibnamefont {Chen}}, \bibinfo {author}
		{\bibfnamefont {Q.}~\bibnamefont {Mao}}, \bibinfo {author} {\bibfnamefont
			{R.}~\bibnamefont {Khan}}, \bibinfo {author} {\bibfnamefont {B.}~\bibnamefont
			{Xu}}, \bibinfo {author} {\bibfnamefont {Y.}~\bibnamefont {Zhou}}, \bibinfo
		{author} {\bibfnamefont {Y.}~\bibnamefont {Zhang}}, \bibinfo {author}
		{\bibfnamefont {J.}~\bibnamefont {Yang}}, \bibinfo {author} {\bibfnamefont
			{B.}~\bibnamefont {Chen}}, \emph {et~al.},\ }\href
	{https://doi.org/10.1007/s11433-016-5798-4} {\bibfield  {journal} {\bibinfo
			{journal} {Science China Physics, Mechanics \& Astronomy}\ }\textbf {\bibinfo
			{volume} {59}},\ \bibinfo {pages} {1} (\bibinfo {year} {2016})}\BibitemShut
	{NoStop}%
	\bibitem [{\citenamefont {Du}\ \emph {et~al.}(2018)\citenamefont {Du},
		\citenamefont {Lou}, \citenamefont {Zhang}, \citenamefont {Zhou},
		\citenamefont {Xu}, \citenamefont {Chen}, \citenamefont {Tang}, \citenamefont
		{Chen}, \citenamefont {Chen}, \citenamefont {Zhu}, \citenamefont {Wang},
		\citenamefont {Yang}, \citenamefont {Wu}, \citenamefont {Yazyev},\ and\
		\citenamefont {Fang}}]{PhysRevB.97.245101}%
	\BibitemOpen
	\bibfield  {author} {\bibinfo {author} {\bibfnamefont {J.}~\bibnamefont
			{Du}}, \bibinfo {author} {\bibfnamefont {Z.}~\bibnamefont {Lou}}, \bibinfo
		{author} {\bibfnamefont {S.}~\bibnamefont {Zhang}}, \bibinfo {author}
		{\bibfnamefont {Y.}~\bibnamefont {Zhou}}, \bibinfo {author} {\bibfnamefont
			{B.}~\bibnamefont {Xu}}, \bibinfo {author} {\bibfnamefont {Q.}~\bibnamefont
			{Chen}}, \bibinfo {author} {\bibfnamefont {Y.}~\bibnamefont {Tang}}, \bibinfo
		{author} {\bibfnamefont {S.}~\bibnamefont {Chen}}, \bibinfo {author}
		{\bibfnamefont {H.}~\bibnamefont {Chen}}, \bibinfo {author} {\bibfnamefont
			{Q.}~\bibnamefont {Zhu}}, \bibinfo {author} {\bibfnamefont {H.}~\bibnamefont
			{Wang}}, \bibinfo {author} {\bibfnamefont {J.}~\bibnamefont {Yang}}, \bibinfo
		{author} {\bibfnamefont {Q.}~\bibnamefont {Wu}}, \bibinfo {author}
		{\bibfnamefont {O.~V.}\ \bibnamefont {Yazyev}},\ and\ \bibinfo {author}
		{\bibfnamefont {M.}~\bibnamefont {Fang}},\ }\href
	{https://doi.org/10.1103/PhysRevB.97.245101} {\bibfield  {journal} {\bibinfo
			{journal} {Phys. Rev. B}\ }\textbf {\bibinfo {volume} {97}},\ \bibinfo
		{pages} {245101} (\bibinfo {year} {2018})}\BibitemShut {NoStop}%
	\bibitem [{\citenamefont {Huang}\ \emph {et~al.}(2015)\citenamefont {Huang},
		\citenamefont {Zhao}, \citenamefont {Long}, \citenamefont {Wang},
		\citenamefont {Chen}, \citenamefont {Yang}, \citenamefont {Liang},
		\citenamefont {Xue}, \citenamefont {Weng}, \citenamefont {Fang},
		\citenamefont {Dai},\ and\ \citenamefont {Chen}}]{PhysRevX.5.031023}%
	\BibitemOpen
	\bibfield  {author} {\bibinfo {author} {\bibfnamefont {X.}~\bibnamefont
			{Huang}}, \bibinfo {author} {\bibfnamefont {L.}~\bibnamefont {Zhao}},
		\bibinfo {author} {\bibfnamefont {Y.}~\bibnamefont {Long}}, \bibinfo {author}
		{\bibfnamefont {P.}~\bibnamefont {Wang}}, \bibinfo {author} {\bibfnamefont
			{D.}~\bibnamefont {Chen}}, \bibinfo {author} {\bibfnamefont {Z.}~\bibnamefont
			{Yang}}, \bibinfo {author} {\bibfnamefont {H.}~\bibnamefont {Liang}},
		\bibinfo {author} {\bibfnamefont {M.}~\bibnamefont {Xue}}, \bibinfo {author}
		{\bibfnamefont {H.}~\bibnamefont {Weng}}, \bibinfo {author} {\bibfnamefont
			{Z.}~\bibnamefont {Fang}}, \bibinfo {author} {\bibfnamefont {X.}~\bibnamefont
			{Dai}},\ and\ \bibinfo {author} {\bibfnamefont {G.}~\bibnamefont {Chen}},\
	}\href {https://doi.org/10.1103/PhysRevX.5.031023} {\bibfield  {journal}
		{\bibinfo  {journal} {Phys. Rev. X}\ }\textbf {\bibinfo {volume} {5}},\
		\bibinfo {pages} {031023} (\bibinfo {year} {2015})}\BibitemShut {NoStop}%
	\bibitem [{\citenamefont {Takatsu}\ \emph {et~al.}(2013)\citenamefont
		{Takatsu}, \citenamefont {Ishikawa}, \citenamefont {Yonezawa}, \citenamefont
		{Yoshino}, \citenamefont {Shishidou}, \citenamefont {Oguchi}, \citenamefont
		{Murata},\ and\ \citenamefont {Maeno}}]{PhysRevLett.111.056601}%
	\BibitemOpen
	\bibfield  {author} {\bibinfo {author} {\bibfnamefont {H.}~\bibnamefont
			{Takatsu}}, \bibinfo {author} {\bibfnamefont {J.~J.}\ \bibnamefont
			{Ishikawa}}, \bibinfo {author} {\bibfnamefont {S.}~\bibnamefont {Yonezawa}},
		\bibinfo {author} {\bibfnamefont {H.}~\bibnamefont {Yoshino}}, \bibinfo
		{author} {\bibfnamefont {T.}~\bibnamefont {Shishidou}}, \bibinfo {author}
		{\bibfnamefont {T.}~\bibnamefont {Oguchi}}, \bibinfo {author} {\bibfnamefont
			{K.}~\bibnamefont {Murata}},\ and\ \bibinfo {author} {\bibfnamefont
			{Y.}~\bibnamefont {Maeno}},\ }\href
	{https://doi.org/10.1103/PhysRevLett.111.056601} {\bibfield  {journal}
		{\bibinfo  {journal} {Phys. Rev. Lett.}\ }\textbf {\bibinfo {volume} {111}},\
		\bibinfo {pages} {056601} (\bibinfo {year} {2013})}\BibitemShut {NoStop}%
	\bibitem [{\citenamefont {Mun}\ \emph {et~al.}(2012)\citenamefont {Mun},
		\citenamefont {Ko}, \citenamefont {Miller}, \citenamefont {Samolyuk},
		\citenamefont {Bud'ko},\ and\ \citenamefont {Canfield}}]{PhysRevB.85.035135}%
	\BibitemOpen
	\bibfield  {author} {\bibinfo {author} {\bibfnamefont {E.}~\bibnamefont
			{Mun}}, \bibinfo {author} {\bibfnamefont {H.}~\bibnamefont {Ko}}, \bibinfo
		{author} {\bibfnamefont {G.~J.}\ \bibnamefont {Miller}}, \bibinfo {author}
		{\bibfnamefont {G.~D.}\ \bibnamefont {Samolyuk}}, \bibinfo {author}
		{\bibfnamefont {S.~L.}\ \bibnamefont {Bud'ko}},\ and\ \bibinfo {author}
		{\bibfnamefont {P.~C.}\ \bibnamefont {Canfield}},\ }\href
	{https://doi.org/10.1103/PhysRevB.85.035135} {\bibfield  {journal} {\bibinfo
			{journal} {Phys. Rev. B}\ }\textbf {\bibinfo {volume} {85}},\ \bibinfo
		{pages} {035135} (\bibinfo {year} {2012})}\BibitemShut {NoStop}%
	\bibitem [{\citenamefont {Chen}\ \emph {et~al.}(2016)\citenamefont {Chen},
		\citenamefont {Lv}, \citenamefont {Luo}, \citenamefont {Lu}, \citenamefont
		{Pei}, \citenamefont {Lin}, \citenamefont {Han}, \citenamefont {Zhu},
		\citenamefont {Song},\ and\ \citenamefont {Sun}}]{PhysRevB.94.235154}%
	\BibitemOpen
	\bibfield  {author} {\bibinfo {author} {\bibfnamefont {F.~C.}\ \bibnamefont
			{Chen}}, \bibinfo {author} {\bibfnamefont {H.~Y.}\ \bibnamefont {Lv}},
		\bibinfo {author} {\bibfnamefont {X.}~\bibnamefont {Luo}}, \bibinfo {author}
		{\bibfnamefont {W.~J.}\ \bibnamefont {Lu}}, \bibinfo {author} {\bibfnamefont
			{Q.~L.}\ \bibnamefont {Pei}}, \bibinfo {author} {\bibfnamefont {G.~T.}\
			\bibnamefont {Lin}}, \bibinfo {author} {\bibfnamefont {Y.~Y.}\ \bibnamefont
			{Han}}, \bibinfo {author} {\bibfnamefont {X.~B.}\ \bibnamefont {Zhu}},
		\bibinfo {author} {\bibfnamefont {W.~H.}\ \bibnamefont {Song}},\ and\
		\bibinfo {author} {\bibfnamefont {Y.~P.}\ \bibnamefont {Sun}},\ }\href
	{https://doi.org/10.1103/PhysRevB.94.235154} {\bibfield  {journal} {\bibinfo
			{journal} {Phys. Rev. B}\ }\textbf {\bibinfo {volume} {94}},\ \bibinfo
		{pages} {235154} (\bibinfo {year} {2016})}\BibitemShut {NoStop}%
	\bibitem [{\citenamefont {Ali}\ \emph {et~al.}(2014)\citenamefont {Ali},
		\citenamefont {Xiong}, \citenamefont {Flynn}, \citenamefont {Tao},
		\citenamefont {Gibson}, \citenamefont {Schoop}, \citenamefont {Liang},
		\citenamefont {Haldolaarachchige}, \citenamefont {Hirschberger},
		\citenamefont {Ong} \emph {et~al.}}]{ali2014large}%
	\BibitemOpen
	\bibfield  {author} {\bibinfo {author} {\bibfnamefont {M.~N.}\ \bibnamefont
			{Ali}}, \bibinfo {author} {\bibfnamefont {J.}~\bibnamefont {Xiong}}, \bibinfo
		{author} {\bibfnamefont {S.}~\bibnamefont {Flynn}}, \bibinfo {author}
		{\bibfnamefont {J.}~\bibnamefont {Tao}}, \bibinfo {author} {\bibfnamefont
			{Q.~D.}\ \bibnamefont {Gibson}}, \bibinfo {author} {\bibfnamefont {L.~M.}\
			\bibnamefont {Schoop}}, \bibinfo {author} {\bibfnamefont {T.}~\bibnamefont
			{Liang}}, \bibinfo {author} {\bibfnamefont {N.}~\bibnamefont
			{Haldolaarachchige}}, \bibinfo {author} {\bibfnamefont {M.}~\bibnamefont
			{Hirschberger}}, \bibinfo {author} {\bibfnamefont {N.~P.}\ \bibnamefont
			{Ong}}, \emph {et~al.},\ }\href {https://doi.org/10.1038/nature13763}
	{\bibfield  {journal} {\bibinfo  {journal} {Nature}\ }\textbf {\bibinfo
			{volume} {514}},\ \bibinfo {pages} {205} (\bibinfo {year}
		{2014})}\BibitemShut {NoStop}%
	\bibitem [{\citenamefont {Chen}\ \emph {et~al.}(2020)\citenamefont {Chen},
		\citenamefont {Lou}, \citenamefont {Zhang}, \citenamefont {Xu}, \citenamefont
		{Zhou}, \citenamefont {Chen}, \citenamefont {Chen}, \citenamefont {Du},
		\citenamefont {Wang}, \citenamefont {Yang}, \citenamefont {Wu}, \citenamefont
		{Yazyev},\ and\ \citenamefont {Fang}}]{PhysRevB.102.165133}%
	\BibitemOpen
	\bibfield  {author} {\bibinfo {author} {\bibfnamefont {Q.}~\bibnamefont
			{Chen}}, \bibinfo {author} {\bibfnamefont {Z.}~\bibnamefont {Lou}}, \bibinfo
		{author} {\bibfnamefont {S.}~\bibnamefont {Zhang}}, \bibinfo {author}
		{\bibfnamefont {B.}~\bibnamefont {Xu}}, \bibinfo {author} {\bibfnamefont
			{Y.}~\bibnamefont {Zhou}}, \bibinfo {author} {\bibfnamefont {H.}~\bibnamefont
			{Chen}}, \bibinfo {author} {\bibfnamefont {S.}~\bibnamefont {Chen}}, \bibinfo
		{author} {\bibfnamefont {J.}~\bibnamefont {Du}}, \bibinfo {author}
		{\bibfnamefont {H.}~\bibnamefont {Wang}}, \bibinfo {author} {\bibfnamefont
			{J.}~\bibnamefont {Yang}}, \bibinfo {author} {\bibfnamefont {Q.}~\bibnamefont
			{Wu}}, \bibinfo {author} {\bibfnamefont {O.~V.}\ \bibnamefont {Yazyev}},\
		and\ \bibinfo {author} {\bibfnamefont {M.}~\bibnamefont {Fang}},\ }\href
	{https://doi.org/10.1103/PhysRevB.102.165133} {\bibfield  {journal} {\bibinfo
			{journal} {Phys. Rev. B}\ }\textbf {\bibinfo {volume} {102}},\ \bibinfo
		{pages} {165133} (\bibinfo {year} {2020})}\BibitemShut {NoStop}%
	\bibitem [{\citenamefont {Zhou}\ \emph {et~al.}(2020)\citenamefont {Zhou},
		\citenamefont {Lou}, \citenamefont {Zhang}, \citenamefont {Chen},
		\citenamefont {Chen}, \citenamefont {Xu}, \citenamefont {Du}, \citenamefont
		{Yang}, \citenamefont {Wang}, \citenamefont {Xi}, \citenamefont {Pi},
		\citenamefont {Wu}, \citenamefont {Yazyev},\ and\ \citenamefont
		{Fang}}]{PhysRevB.102.115145}%
	\BibitemOpen
	\bibfield  {author} {\bibinfo {author} {\bibfnamefont {Y.}~\bibnamefont
			{Zhou}}, \bibinfo {author} {\bibfnamefont {Z.}~\bibnamefont {Lou}}, \bibinfo
		{author} {\bibfnamefont {S.}~\bibnamefont {Zhang}}, \bibinfo {author}
		{\bibfnamefont {H.}~\bibnamefont {Chen}}, \bibinfo {author} {\bibfnamefont
			{Q.}~\bibnamefont {Chen}}, \bibinfo {author} {\bibfnamefont {B.}~\bibnamefont
			{Xu}}, \bibinfo {author} {\bibfnamefont {J.}~\bibnamefont {Du}}, \bibinfo
		{author} {\bibfnamefont {J.}~\bibnamefont {Yang}}, \bibinfo {author}
		{\bibfnamefont {H.}~\bibnamefont {Wang}}, \bibinfo {author} {\bibfnamefont
			{C.}~\bibnamefont {Xi}}, \bibinfo {author} {\bibfnamefont {L.}~\bibnamefont
			{Pi}}, \bibinfo {author} {\bibfnamefont {Q.}~\bibnamefont {Wu}}, \bibinfo
		{author} {\bibfnamefont {O.~V.}\ \bibnamefont {Yazyev}},\ and\ \bibinfo
		{author} {\bibfnamefont {M.}~\bibnamefont {Fang}},\ }\href
	{https://doi.org/10.1103/PhysRevB.102.115145} {\bibfield  {journal} {\bibinfo
			{journal} {Phys. Rev. B}\ }\textbf {\bibinfo {volume} {102}},\ \bibinfo
		{pages} {115145} (\bibinfo {year} {2020})}\BibitemShut {NoStop}%
	\bibitem [{\citenamefont {Wang}\ \emph {et~al.}(2015)\citenamefont {Wang},
		\citenamefont {Thoutam}, \citenamefont {Xiao}, \citenamefont {Hu},
		\citenamefont {Das}, \citenamefont {Mao}, \citenamefont {Wei}, \citenamefont
		{Divan}, \citenamefont {Luican-Mayer}, \citenamefont {Crabtree},\ and\
		\citenamefont {Kwok}}]{PhysRevB.92.180402}%
	\BibitemOpen
	\bibfield  {author} {\bibinfo {author} {\bibfnamefont {Y.~L.}\ \bibnamefont
			{Wang}}, \bibinfo {author} {\bibfnamefont {L.~R.}\ \bibnamefont {Thoutam}},
		\bibinfo {author} {\bibfnamefont {Z.~L.}\ \bibnamefont {Xiao}}, \bibinfo
		{author} {\bibfnamefont {J.}~\bibnamefont {Hu}}, \bibinfo {author}
		{\bibfnamefont {S.}~\bibnamefont {Das}}, \bibinfo {author} {\bibfnamefont
			{Z.~Q.}\ \bibnamefont {Mao}}, \bibinfo {author} {\bibfnamefont
			{J.}~\bibnamefont {Wei}}, \bibinfo {author} {\bibfnamefont {R.}~\bibnamefont
			{Divan}}, \bibinfo {author} {\bibfnamefont {A.}~\bibnamefont {Luican-Mayer}},
		\bibinfo {author} {\bibfnamefont {G.~W.}\ \bibnamefont {Crabtree}},\ and\
		\bibinfo {author} {\bibfnamefont {W.~K.}\ \bibnamefont {Kwok}},\ }\href
	{https://doi.org/10.1103/PhysRevB.92.180402} {\bibfield  {journal} {\bibinfo
			{journal} {Phys. Rev. B}\ }\textbf {\bibinfo {volume} {92}},\ \bibinfo
		{pages} {180402} (\bibinfo {year} {2015})}\BibitemShut {NoStop}%
	\bibitem [{\citenamefont {Bai}\ \emph {et~al.}(2024)\citenamefont {Bai},
		\citenamefont {Xiang}, \citenamefont {Pan}, \citenamefont {Zhang},
		\citenamefont {Chen}, \citenamefont {Han}, \citenamefont {Xu},\ and\
		\citenamefont {Xu}}]{bai2024non}%
	\BibitemOpen
	\bibfield  {author} {\bibinfo {author} {\bibfnamefont {Y.}~\bibnamefont
			{Bai}}, \bibinfo {author} {\bibfnamefont {X.}~\bibnamefont {Xiang}}, \bibinfo
		{author} {\bibfnamefont {S.}~\bibnamefont {Pan}}, \bibinfo {author}
		{\bibfnamefont {S.}~\bibnamefont {Zhang}}, \bibinfo {author} {\bibfnamefont
			{H.~C.~X.}\ \bibnamefont {Chen}}, \bibinfo {author} {\bibfnamefont
			{Z.}~\bibnamefont {Han}}, \bibinfo {author} {\bibfnamefont {G.}~\bibnamefont
			{Xu}},\ and\ \bibinfo {author} {\bibfnamefont {F.}~\bibnamefont {Xu}},\
	}\href {https://arxiv.org/abs/2409.14855} {} (\bibinfo {year} {2024}),\
	\Eprint {https://arxiv.org/abs/2409.14855} {arXiv:2409.14855
		[cond-mat.mtrl-sci]} \BibitemShut {NoStop}%
	\bibitem [{\citenamefont {Xu}\ \emph {et~al.}(2021)\citenamefont {Xu},
		\citenamefont {Han}, \citenamefont {Wang}, \citenamefont {Thoutam},
		\citenamefont {Pate}, \citenamefont {Li}, \citenamefont {Zhang},
		\citenamefont {Wang}, \citenamefont {Fotovat}, \citenamefont {Welp},
		\citenamefont {Zhou}, \citenamefont {Kwok}, \citenamefont {Chung},
		\citenamefont {Kanatzidis},\ and\ \citenamefont {Xiao}}]{PhysRevX.11.041029}%
	\BibitemOpen
	\bibfield  {author} {\bibinfo {author} {\bibfnamefont {J.}~\bibnamefont
			{Xu}}, \bibinfo {author} {\bibfnamefont {F.}~\bibnamefont {Han}}, \bibinfo
		{author} {\bibfnamefont {T.-T.}\ \bibnamefont {Wang}}, \bibinfo {author}
		{\bibfnamefont {L.~R.}\ \bibnamefont {Thoutam}}, \bibinfo {author}
		{\bibfnamefont {S.~E.}\ \bibnamefont {Pate}}, \bibinfo {author}
		{\bibfnamefont {M.}~\bibnamefont {Li}}, \bibinfo {author} {\bibfnamefont
			{X.}~\bibnamefont {Zhang}}, \bibinfo {author} {\bibfnamefont {Y.-L.}\
			\bibnamefont {Wang}}, \bibinfo {author} {\bibfnamefont {R.}~\bibnamefont
			{Fotovat}}, \bibinfo {author} {\bibfnamefont {U.}~\bibnamefont {Welp}},
		\bibinfo {author} {\bibfnamefont {X.}~\bibnamefont {Zhou}}, \bibinfo {author}
		{\bibfnamefont {W.-K.}\ \bibnamefont {Kwok}}, \bibinfo {author}
		{\bibfnamefont {D.~Y.}\ \bibnamefont {Chung}}, \bibinfo {author}
		{\bibfnamefont {M.~G.}\ \bibnamefont {Kanatzidis}},\ and\ \bibinfo {author}
		{\bibfnamefont {Z.-L.}\ \bibnamefont {Xiao}},\ }\href
	{https://doi.org/10.1103/PhysRevX.11.041029} {\bibfield  {journal} {\bibinfo
			{journal} {Phys. Rev. X}\ }\textbf {\bibinfo {volume} {11}},\ \bibinfo
		{pages} {041029} (\bibinfo {year} {2021})}\BibitemShut {NoStop}%
	\bibitem [{\citenamefont {Zhang}\ \emph {et~al.}(2024)\citenamefont {Zhang},
		\citenamefont {Liu}, \citenamefont {Pi}, \citenamefont {Fang}, \citenamefont
		{Weng},\ and\ \citenamefont {Wu}}]{PhysRevB.110.205132}%
	\BibitemOpen
	\bibfield  {author} {\bibinfo {author} {\bibfnamefont {S.}~\bibnamefont
			{Zhang}}, \bibinfo {author} {\bibfnamefont {Z.}~\bibnamefont {Liu}}, \bibinfo
		{author} {\bibfnamefont {H.}~\bibnamefont {Pi}}, \bibinfo {author}
		{\bibfnamefont {Z.}~\bibnamefont {Fang}}, \bibinfo {author} {\bibfnamefont
			{H.}~\bibnamefont {Weng}},\ and\ \bibinfo {author} {\bibfnamefont
			{Q.}~\bibnamefont {Wu}},\ }\href
	{https://doi.org/10.1103/PhysRevB.110.205132} {\bibfield  {journal} {\bibinfo
			{journal} {Phys. Rev. B}\ }\textbf {\bibinfo {volume} {110}},\ \bibinfo
		{pages} {205132} (\bibinfo {year} {2024})}\BibitemShut {NoStop}%
	\bibitem [{\citenamefont {SupplementalMaterial}(2024)}]{SupplementalMaterial}%
	\BibitemOpen
	\bibfield  {author} {\bibinfo {author} {\bibnamefont {SupplementalMaterial}}}
	(\bibinfo {year} {2024})\BibitemShut {NoStop}%
	\bibitem [{\citenamefont {Urata}\ \emph {et~al.}(2024)\citenamefont {Urata},
		\citenamefont {Hattori},\ and\ \citenamefont {Ikuta}}]{urata2024high}%
	\BibitemOpen
	\bibfield  {author} {\bibinfo {author} {\bibfnamefont {T.}~\bibnamefont
			{Urata}}, \bibinfo {author} {\bibfnamefont {W.}~\bibnamefont {Hattori}},\
		and\ \bibinfo {author} {\bibfnamefont {H.}~\bibnamefont {Ikuta}},\ }\href
	{https://doi.org/10.1103/PhysRevMaterials.8.084412} {\bibfield  {journal}
		{\bibinfo  {journal} {Phys. Rev. Mater.}\ }\textbf {\bibinfo {volume} {8}},\
		\bibinfo {pages} {084412} (\bibinfo {year} {2024})}\BibitemShut {NoStop}%
	\bibitem [{\citenamefont {{\v{S}}mejkal}\ \emph {et~al.}(2020)\citenamefont
		{{\v{S}}mejkal}, \citenamefont {Gonz{\'a}lez-Hern{\'a}ndez}, \citenamefont
		{Jungwirth},\ and\ \citenamefont {Sinova}}]{vsmejkal2020crystal}%
	\BibitemOpen
	\bibfield  {author} {\bibinfo {author} {\bibfnamefont {L.}~\bibnamefont
			{{\v{S}}mejkal}}, \bibinfo {author} {\bibfnamefont {R.}~\bibnamefont
			{Gonz{\'a}lez-Hern{\'a}ndez}}, \bibinfo {author} {\bibfnamefont
			{T.}~\bibnamefont {Jungwirth}},\ and\ \bibinfo {author} {\bibfnamefont
			{J.}~\bibnamefont {Sinova}},\ }\href@noop {} {\bibfield  {journal} {\bibinfo
			{journal} {Sci. adv.}\ }\textbf {\bibinfo {volume} {6}},\ \bibinfo {pages}
		{eaaz8809} (\bibinfo {year} {2020})}\BibitemShut {NoStop}%
	\bibitem [{\citenamefont {Zhou}\ \emph {et~al.}(2024)\citenamefont {Zhou},
		\citenamefont {Cheng}, \citenamefont {Hu}, \citenamefont {Liu}, \citenamefont
		{Pan},\ and\ \citenamefont {Song}}]{zhou2024crystal}%
	\BibitemOpen
	\bibfield  {author} {\bibinfo {author} {\bibfnamefont {Z.}~\bibnamefont
			{Zhou}}, \bibinfo {author} {\bibfnamefont {X.}~\bibnamefont {Cheng}},
		\bibinfo {author} {\bibfnamefont {M.}~\bibnamefont {Hu}}, \bibinfo {author}
		{\bibfnamefont {J.}~\bibnamefont {Liu}}, \bibinfo {author} {\bibfnamefont
			{F.}~\bibnamefont {Pan}},\ and\ \bibinfo {author} {\bibfnamefont
			{C.}~\bibnamefont {Song}},\ }\href {https://arxiv.org/abs/2403.07396} {}
	(\bibinfo {year} {2024}),\ \Eprint {https://arxiv.org/abs/2403.07396}
	{arXiv:2403.07396 [cond-mat.mtrl-sci]} \BibitemShut {NoStop}%
	\bibitem [{\citenamefont {Dzyaloshinskii}\ \emph {et~al.}(1957)\citenamefont
		{Dzyaloshinskii} \emph {et~al.}}]{dzyaloshinskii1957}%
	\BibitemOpen
	\bibfield  {author} {\bibinfo {author} {\bibfnamefont {I.}~\bibnamefont
			{Dzyaloshinskii}} \emph {et~al.},\ }\href@noop {} {\bibfield  {journal}
		{\bibinfo  {journal} {Sov. Phys. JETP}\ }\textbf {\bibinfo {volume} {5}},\
		\bibinfo {pages} {1259} (\bibinfo {year} {1957})}\BibitemShut {NoStop}%
\end{thebibliography}
%

\end{document}